\documentclass[iop]{emulateapj}		
\usepackage{epsfig}			
\usepackage{graphicx,color}		
\usepackage{amssymb}			
\usepackage{color}			
\usepackage{url}			
\usepackage{amsmath}			
\usepackage{rotating}			
\usepackage{float}			
\usepackage{textcomp}			
\usepackage{psfig}
\usepackage{dcolumn}
\usepackage{times}
\usepackage{tabularx}
\usepackage[colorlinks=true,citecolor=blue]{hyperref}
\usepackage[english]{babel}

\shorttitle{The effects of transients on photospheric and chromospheric power distributions}
\shortauthors{T. Samanta et al.}

\begin{document}

\title{The effects of transients on photospheric and chromospheric power distributions}

\author{T. Samanta$^{1}$,
V. M. J. Henriques$^{2}$,
D. Banerjee$^{1,3}$,
S. Krishna Prasad$^{2}$,
M. Mathioudakis$^{2}$,
D. Jess$^{2}$,
V. Pant$^{1}$
}
 
\affil{$^{1}$Indian Institute of Astrophysics, Koramangala, Bangalore 560034, India. e-mail: {\color{blue}{tsamanta@iiap.res.in}}\\
$^{2}$Astrophysics Research Centre, School of Mathematics and Physics, Queen's University Belfast, Belfast BT7 1NN, UK. e-mail: {\color{blue}{v.henriques@qub.ac.uk}}\\
$^{3}$Center of Excellence in Space Sciences, IISER Kolkata, India. e-mail: {\color{blue}{dipu@iiap.res.in}}
}
  
\begin{abstract}
We have observed a quiet Sun region with the Swedish 1-meter Solar Telescope (SST) equipped with CRISP Imaging SpectroPolarimeter. High-resolution, 
high-cadence, H$\alpha$ line scanning images were taken to observe different layers of  the solar atmosphere from the photosphere to upper chromosphere. 
We study the distribution of power in different period-bands at different heights. Power maps of the upper photosphere and the lower chromosphere show 
suppressed power surrounding the magnetic-network elements, known as ``magnetic shadows''. These also show enhanced power close to the photosphere, 
traditionally referred to as ``power halos''. 
The interaction between acoustic waves and inclined magnetic fields is generally believed to be responsible for these two effects. 
In this study we explore if small-scale transients can influence the distribution of power at different heights. We show that the presence of transients, 
like mottles, Rapid Blueshifted Excursions (RBEs) and Rapid Redshifted Excursions (RREs), can strongly influence the power-maps.
The short and finite lifetime
of these events strongly affects all powermaps, potentially influencing the observed power
distribution. We show that Doppler-shifted transients like RBEs and RREs  
that occur ubiquitously, can have a dominant effect on the formation of the power halos in the quiet Sun.   
For magnetic shadows, transients like mottles do not seem to have a significant effect in the 
power suppression around 3 minutes and wave interaction may play a key role here.
 Our high cadence observations 
reveal that flows, waves and shocks manifest in presence of magnetic fields 
to form a non-linear magnetohydrodynamic system.

\end{abstract}


\section{Introduction}
\label{introduction}
The solar chromosphere is a layer above the visible solar surface spanning over approximately a thousand kilometers in height.
It plays an important role in understanding the interaction between the relatively cool photospheric plasma and the hot multi-million degree corona.
Small scale magnetic flux concentrations at the boundaries of supergranular cells extend upwards into the chromosphere. 
These flux tubes expand into funnel-like structures with height due to a decrease in the ambient gas pressure.
Some field lines locally connect within the photosphere and produce a canopy-like structure in the chromosphere and some of them reach to corona. 
The chromosphere is still a poorly-understood layer where flows, waves and shocks manifest in the presence of magnetic fields to form an often nonlinear 
magneto-hydrodynamic system.


Waves in the solar atmosphere are studied with great interest as  they carry mechanical energy and also provide insight of the physical parameters 
through seismology \citep{2000SoPh..193..139R,2007SoPh..246....3B,2009SSRv..149..355Z,2012RSPTA.370.3193D,2015SSRv..190..103J}. 
Oscillations are observed ubiquitously throughout the solar atmosphere and are often interpreted in terms of various magnetohydrodynamic (MHD) modes. 
Acoustic waves (p-modes) which are generated inside the Sun, are generally trapped inside it. These waves can freely propagate from the surface into the atmosphere if they have periods shorter than 3.2 minutes (5.2 mHz) which is known as the acoustic cut-off period. The longer periods generally do not propagate to greater heights and are, instead, evanescent.  There is substantial observational evidence for the presence of long period oscillations \citep{2007A&A...461L...1V,2010A&A...510A..41K,2014A&A...567A..62K,2014MNRAS.443.1267B} in the chromosphere  around network magnetic elements. It appears that the presence of strong network magnetic field  changes the scenario \citep{2002ApJ...564..508R,2003ApJ...599..626B}. \citet{1983SoPh...87...77R}, \citet{2006ApJ...640.1153C} and \citet{2008ApJ...676L..85K} argue that strong magnetic fields change the radiative relaxation time, which can increase the cut-off period significantly. There are also suggestions that the field inclination 
plays a very important role in long-period wave propagation \citep{2006RSPTA.364..395C,2013ApJ...779..168J,2014A&A...567A..62K}. \citet{2011ApJ...743..142H} show that the field 
inclination is much more important for long-period wave propagation than the radiative relaxation time effect. Highly inclined magnetic fields significantly increase the cut-off period and create magneto-acoustic portals \citep{2006ApJ...648L.151J} for the propagation of long-period waves in the chromosphere. This is commonly referred to as leakage of photospheric oscillations into the chromosphere \citep{2004Natur.430..536D}. The ``leakage" of long-period photospheric oscillations takes place through magnetic network elements through restricted areas. Recent studies show that a good 
fraction of power is present above the cut-off period at higher layers around the quiet magnetic network elements \citep{2001ApJ...554..424J,2003A&A...405..769M,2007A&A...471..961M,2007A&A...461L...1V,2010A&A...524A..12K}. Two-dimensional power maps of period bands around 3 minutes, reveal two distinct phenomena above network and around elements. 
One is known as ''power halos'', which are upper-photospheric regions where the wave power is enhanced. The other
is ''magnetic shadows'', which refers to the regions of the power suppression around network
elements in the chromosphere.


Many researchers have suggested that the interaction between the acoustic waves and the magnetic fields is responsible for the formation of magnetic shadows and power halos \citep{2001ApJ...554..424J,2003A&A...405..769M,2007A&A...471..961M,2007A&A...461L...1V,2010A&A...524A..12K}. It was proposed that the upward propagating acoustic waves change their nature at the magnetic canopy, a layer where the gas pressure becomes equal to the magnetic pressure, and undergo mode conversion and transmission processes \citep{2010AN....331..915N}. Simulations show that acoustic waves generally transfer their energy partly to the slow magneto-acoustic waves (mode transmission) and partly to the fast magneto-acoustic waves (mode conversion) at the canopy \citep{2012A&A...538A..79N}. Due to high velocity gradients, the fast mode undergoes reflection at the canopy and increases the oscillation power at lower heights, creating power halos \citep{2012A&A...542L..30N}. In contrast, the slow mode continues to propagate 
along the 
slanted 
magnetic field lines. \citet{2014A&A...567A..62K} argue that the key parameter in mode conversion mechanism is the attack angle (the angle between the direction of wave propagation and the magnetic field) and the period of the acoustic waves. They show that the transmission is generally favored at small attack angles and long periods, while the conversion dominates when the period is 
small and the attack angle is large, which causes the power halos and magnetic shadows to form around magnetic network regions.

However, most of the earlier studies that put forward the theory based on magneto acoustic-wave 
reflection did not  consider the effect of  transients,  nor the evolution of magnetic fields and 
other factors leading to changes in the visible chromospheric canopy. 
Earlier observations
were also limited by low temporal resolution to study the influence of short-lived transients in
detail.
With our high spatial and temporal resolution observations taken with the 1-m Swedish Solar Telescope  we revisit the subject and attempt to provide an alternative  interpretation for the formation of the magnetic shadow and power halo. Section 2 describes the observations along with data reduction procedures. Section 3 provides results in terms of power distribution at several heights in the chromosphere. Section 4 deals with possible scenarios which can explain our observations and compares with earlier interpretations. Finally, conclusions are drawn in Section 5.

\section{Observations and Data Processing}
\begin{figure}
\centering
\includegraphics[angle=90,trim = 5mm 42.9mm 12mm 29.5mm, clip,width=8.0cm]{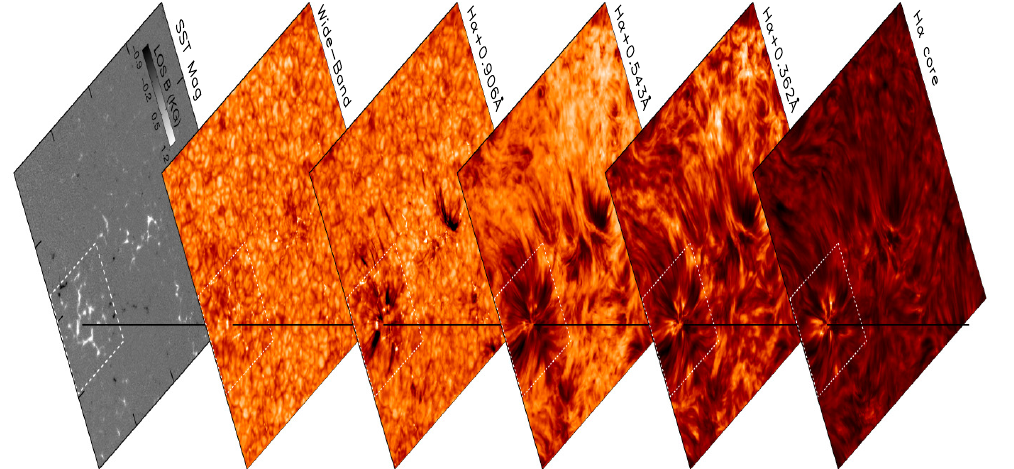}
\caption{Images of a quiet region as seen in different layers of the solar atmosphere along with the corresponding magnetogram from photosphere at the bottom.
\textit{Bottom to Top}: Line-of-sight (LOS) magnetogram obtained by using Fe 6302~\r{A} Stokes V profiles, visible continuum, and narrow band filter images 
taken at different positions across the H$\alpha$ line profile 
as indicated (H$\alpha$~+~0.906~\r{A}, H$\alpha$~+~0.543~\r{A}, H$\alpha$~+~0.362~\r{A} and  H$\alpha$ core). 
The long tick marks on the magnetogram represent 10 Mm intervals. The region outline by the dotted line covers a network region is further studied in Figure~\ref{xt_s6_cut1}, \ref{xt_s3_cut1} and \ref{core_dop}.
An animation of this figure is available online.}
\label{observation} 
\end{figure}
\begin{figure*}[ht]
\centering
\includegraphics[angle=90,trim = 5mm 42.9mm 3mm 29.4mm, clip,width=6.0cm]{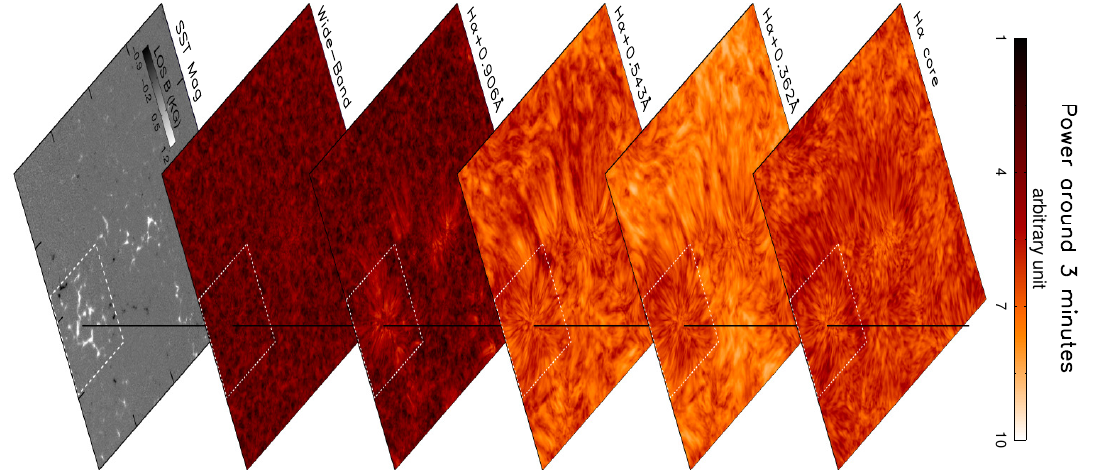}\includegraphics[angle=90,trim = 5mm 42.9mm 3mm 29.4mm, clip,width=6.0cm]{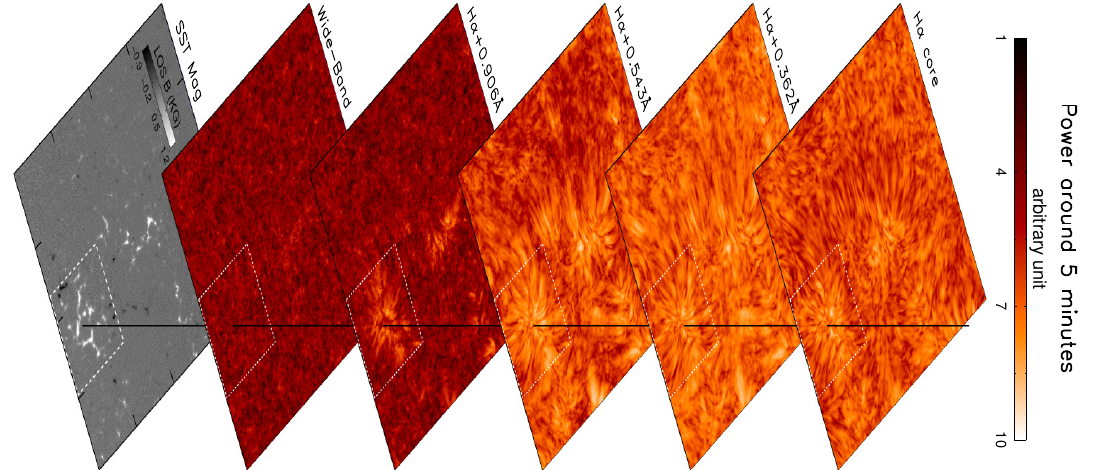}\includegraphics[angle=90,trim = 5mm 42.9mm 3mm 29.4mm, clip,width=6.0cm]{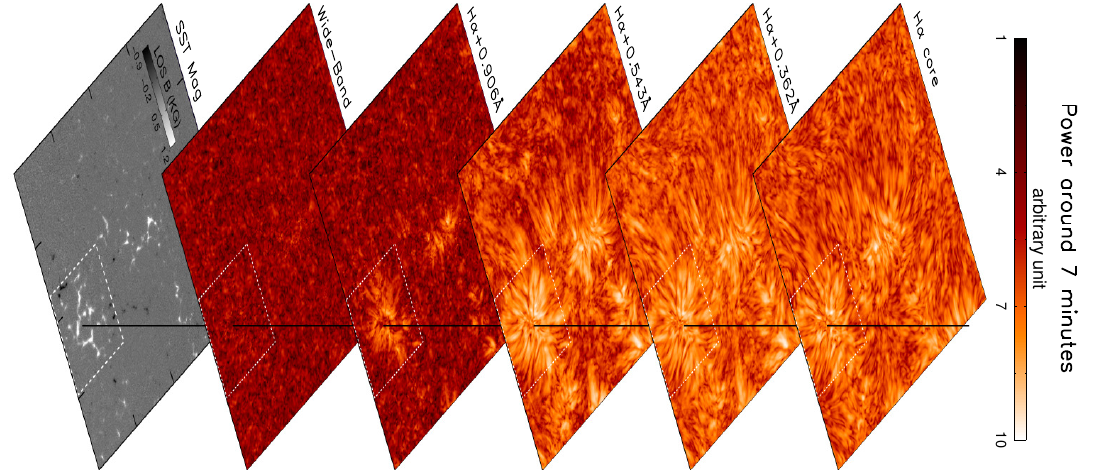}
\caption{Power-maps in different layers in three 1-minute wide period bands around 3, 5 and 7 minutes. Corresponding photospheric magnetograms are shown at the bottom. The long tick marks on the magnetogram represent 10 Mm intervals.}
\label{power_map} 
\end{figure*}
Observations of a quiet-Sun region were made on 2013 May 3, from 09:06 UT to 09:35 UT using the CRisp Imaging SpectroPolarimeter \citep[CRISP;][]{2006A&A...447.1111S, 2008ApJ...689L..69S}, at the 1-m Swedish Solar Telescope \citep[SST;][]{2003SPIE.4853..341S}. Images were taken at 7 wavelength positions scanning through the H$\alpha$ line, at -0.906 {\AA}, -0.543 {\AA}, -0.362 {\AA}, 0.000 {\AA}, 0.362 {\AA}, 0.543 {\AA}, and +0.906 {\AA} from line core corresponding to a velocity range of -41 to +41 km s$^{-1}$.
Adaptive optics was employed in the observations with the upgraded 85-electrode system \citep{2003SPIE.4853..370S}. 

All the data were reconstructed using Multi-Object Multi-Frame 
Blind Deconvolution \citep[MOMFBD;][]{2002SPIE.4792..146L,2005SoPh..228..191V}, with 51~Karhunen-Lo\`{e}ve modes sorted by order 
of atmospheric significance and $88\times88$ pixel subfields. 
An early version of the pipeline described in \citet{2015A&A...573A..40D} was used.
Destretching \citep{1994ApJ...430..413S} was used together with auxiliary wide-band objects for 
consistent co-alignment of different narrow-band passbands, as described in \citet{2012A&A...548A.114H}.
Spatial sampling is 0$''$.058 pixel$^{-1}$, and the spatial resolution reaches up to 0$''$.16 in H$\alpha$ covering a field-of-view (FOV) of 40$\times$40 Mm$^{2}$. 
After reconstruction, the cadence of a full spectral scan was 1.34 s. In this work, we also made use of wide-band images obtained with the CRISP reference camera. 
This camera is behind the H$\alpha$ pre-filter but before the double Fabry-P\'erot. The pre-filter has a 1~nm passband centered at the core of the line. The images from this camera provide the 
anchor channel for MOMFBD reconstruction and the reference for all post-reconstruction destretch-based techniques. 
The vast majority of the light contributing to the images from this camera come from the photospheric wings of the H$\alpha$ line.
The cadence of wide-band was also 1.34 s.

Line-of-Sight (LOS) magnetograms were produced 
from the Stokes V output of Fe 6301~{\AA} spectral scans
(taken at 16 wavelength positions) using the centre of gravity (COG) method \citep{1979A&A....74....1R, 2003ApJ...592.1225U}. 
These scans were acquired at a cadence of $\sim$5 minutes  over the same FOV. 
The same H$\alpha$ camera was used for obtaining Stokes V data. Hence there were gaps of $\sim$27~s at $\sim$5, 11, 16 and 21 minutes of observation.
We have interpolated these data gaps using a spline function to obtain a 
regular cadence for time series analysis. 
Note that the treatment via spline-fitting is smoothly 'bridging' intensity in the time-series whereas RBE/RREs and mottles cause strong dips in the intensity. 
Further details on the observations and data reduction are given in \cite{2015ApJ...802...26K} and \cite{2016arXiv160204820H}. 

H$\alpha$ core maps were produced using Doppler compensation. For this, at first, we increased the line profile sampling by a factor of 10 times more than the original using spline interpolation, 
then the minimum value of the profile is calculated at each pixel to produce the Doppler-compensated H$\alpha$ core maps. 
This procedure minimizes the effects of strong flows that might shift the position of the line core, 
and thus best represents the emission coming from the line-forming region (see, e.g., \citet{2010ApJ...719L.134J}). The LOS Doppler velocity maps were determined by the COG method.

The H$\alpha$ line core forms at the chromosphere and the wings form at lower atmospheric heights \citep{2006A&A...449.1209L,2012ApJ...749..136L}. 
Filtergram images taken at different positions of the H$\alpha$ line, sample, on average, different atmospheric layers  and are shown in Figure~\ref{observation}. A time lapsed movie of this Figure is also available online. The movie clearly shows the presence of transients. 

\begin{figure}
\centering
\includegraphics[angle=90,trim = 5mm 42.9mm 3mm 29.4mm, clip,width=7.0cm]{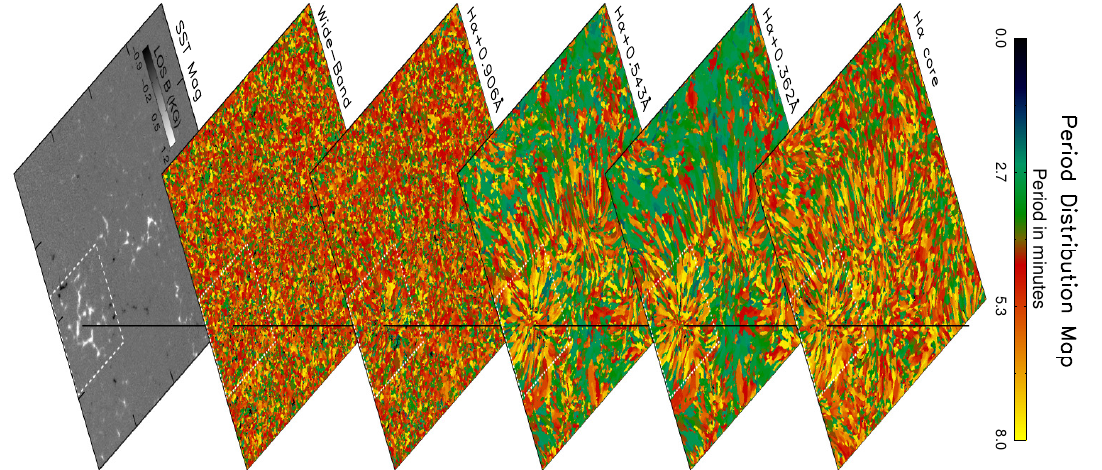} 
\caption{Distribution of dominant periods in different layers along with the corresponding magnetogram at the bottom. The green, red and yellow colors roughly represent periods around 3, 5 and 7 minutes respectively.}
\label{period_map} 
\end{figure}
\begin{figure*}
\centering
\includegraphics[angle=90,clip,width=18cm]{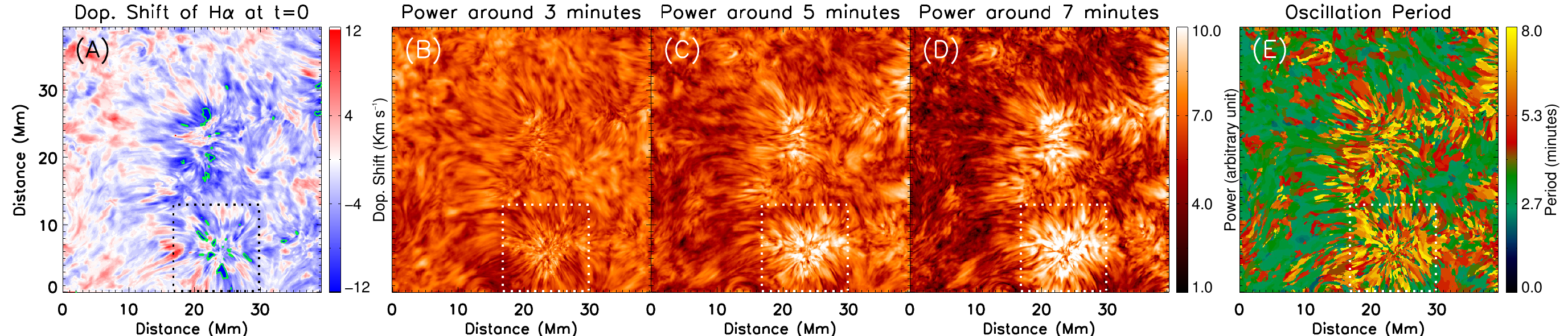}
\caption{(A): H$\alpha$ Doppler shift map (at t=0) obtained using the Center-of-Gravity (COG) method. The  display scale is saturated to $\pm$12 km s$^{-1}$ for better view. The value inside the green contours correspond to higher than  $\pm$12 km s$^{-1}$.  The power-maps at different period bands are shown in panels (B) - (D). (E): Distribution of the dominant periods.}
\label{doppler} 
\end{figure*}
\section{Results}
\subsection{Spatially resolved power distributions in different period bands}
\label{power_period}
\begin{figure}
\centering
\includegraphics[angle=90,clip,width=8cm]{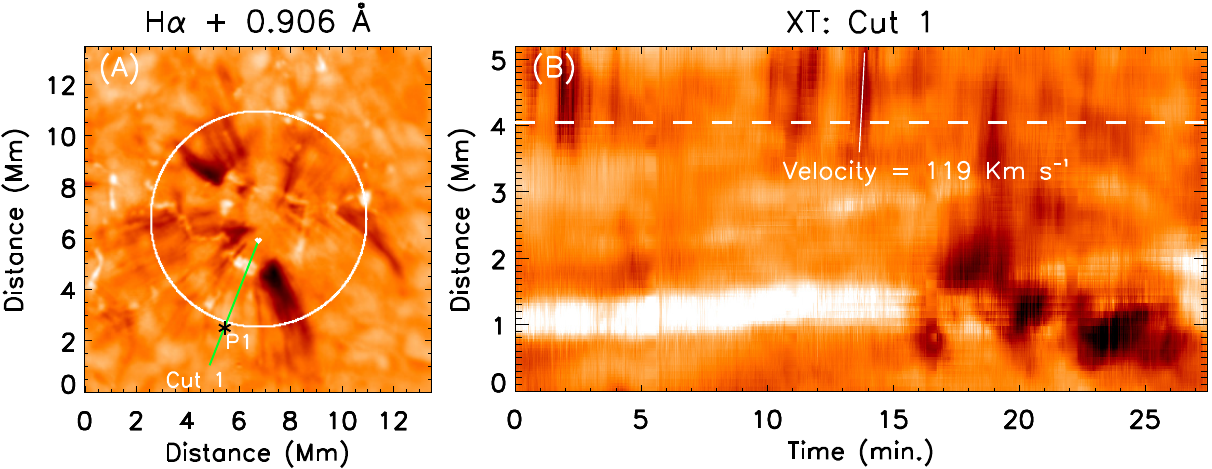}
\includegraphics[angle=90,clip,width=8.5cm]{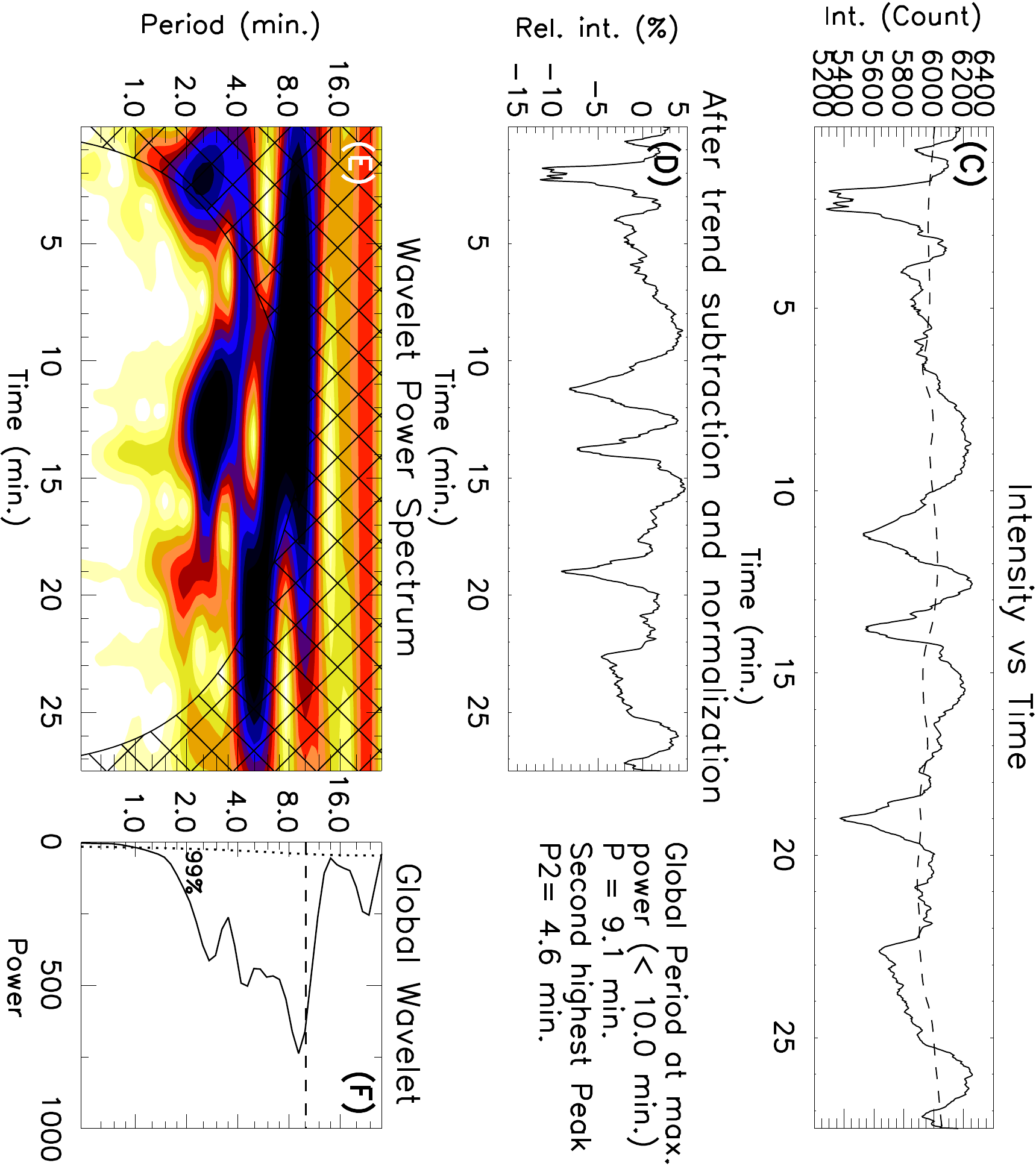}
\caption{(A): H$\alpha$ + 0.906~\AA\ image. A cut along the green line is taken to produce the space-time map shown in (B). The white dot represents the starting point. (B): Temporal evolution along the green line shown in (A).
The dashed line corresponds to the position P1 marked by an asterisk in (A). The slanted solid line indicates the track used for measuring the propagation speed. (C): The intensity variation along the dashed line in (B). The overplotted dashed line represents the background trend. (D): Relative intensity variation after trend subtracttion and normalization. (E): The wavelet power spectrum of the normalized timeseries. The overplotted cross-hatched region is the Cone-Of-Influence
(COI) with darker color representing higher power. (F): Global wavelet power spectrum. The maximum measurable period, 10 minutes (due to COI) 
is shown by a horizontal dashed line. The dotted curve shows the 99\% significance level. 
The two most significant periods identified from the global wavelet power spectrum are printed on top of the global wavelet plot.
An animation of panel A for a bigger field-of-view and also for H$\alpha$ blue wing (H$\alpha$ - 0.906~\r{A}) is available online.}
\label{xt_s6_cut1} 
\end{figure}

We investigate the oscillation properties of the different layers by constructing power-maps. The construction of the power maps was preceded by the removal 
of a background trend from each lightcurve  to obtain the relative percentage intensity variations ($I_{R}$) given by $I_{R}=(I-I_{bg})*{I_{bg}}^{-1}*100$, 
where I is the original intensity and I$_{bg}$ is the background trend. 
The background trend,  I$_{bg}$, is computed from the original lightcurve over a 600 s running average, which when subtracted from the original time series allows intensity fluctuations shorter than 10 minutes to be more readily identified.
The resultant lightcurves are then subjected to wavelet analysis \citep{1998BAMS...79...61T} and 
the global wavelet power spectrum is calculated at each pixel. An example of the computed relative intensity 
variations and the corresponding wavelet analysis results at a single pixel are  shown in Figure~\ref{xt_s6_cut1}. Power-maps were constructed for 3, 5, and 7 minutes period from the global wavelet power spectrum by averaging the power in a one minute bands around each period. Figure~\ref{power_map} 
displays 
these maps stacked in ascending order of atmospheric height for each band. A co-spatial photospheric magnetogram is also shown at the bottom panel for comparison. 
Figure~\ref{power_map} reveals that power is suppressed in rosettes over the network in the 3-minute band at lower chromosphere (H$\alpha+$0.543 and H$\alpha+$ 0.362 \r{A}) and enhanced 
close to the photosphere (H$\alpha+$0.906~\AA) in all the bands. These phenomena are known 
as magnetic shadows \citep{2001ApJ...554..424J,2003A&A...405..769M,2007A&A...471..961M,2007A&A...461L...1V,2010A&A...524A..12K,2014A&A...567A..62K} and power halos \citep{2010A&A...510A..41K}, respectively.

We also make period maps to study the spatial distribution of dominant periods in each layer. The period at maximum power above the 99\% significance level 
is taken as the dominant period at each pixel to construct these maps. The significance levels are calculated assuming white noise \citep{1998BAMS...79...61T}. 
The period distribution maps are shown in Figure~\ref{period_map} along with the photospheric magnetogram.  As in Figure~\ref{power_map} and \ref{period_map}, only the maps 
produced from the red wings of H$\alpha$ are displayed since the blue-wing maps look very similar.
It is evident from the figure, that in the layers dominated by the photosphere (wide-band and H$\alpha+$0.906~\r{A}), 
the well known 5-minute photospheric p-mode oscillation is dominant. At larger heights (H$\alpha+$0.543 and H$\alpha+$ 0.362 \r{A}), 
the 3-minute period becomes dominant for most of the field-of-view, with the exception of the neighbourhood of the network magnetic element
where the longer (5 to 7-minute) periods become dominant. 
The distribution of period (see Figure~\ref{doppler}~E) in the H$\alpha$ Doppler velocity maps show that the 3-minute oscillations cover a  wider extent than that in the corresponding period-maps computed from  H$\alpha$-core intensity (see Figure~\ref{period_map}). 
This behavior was observed earlier by \citet{2007ASPC..368...65D}. The velocity power-maps at different period bands are also shown in Figure~\ref{doppler}~(B-D). It shows 
enhanced power in the higher period-bands around network and suppressed power at lower period band (3-minute) at the same region.
Power/period maps were also generated using fast Fourier transform (FFT) techniques. No significant differences were found when compared to our wavelet results, and hence to avoid duplication we do not include these figures here.

\begin{figure}
\centering
\includegraphics[angle=90,clip,width=8cm]{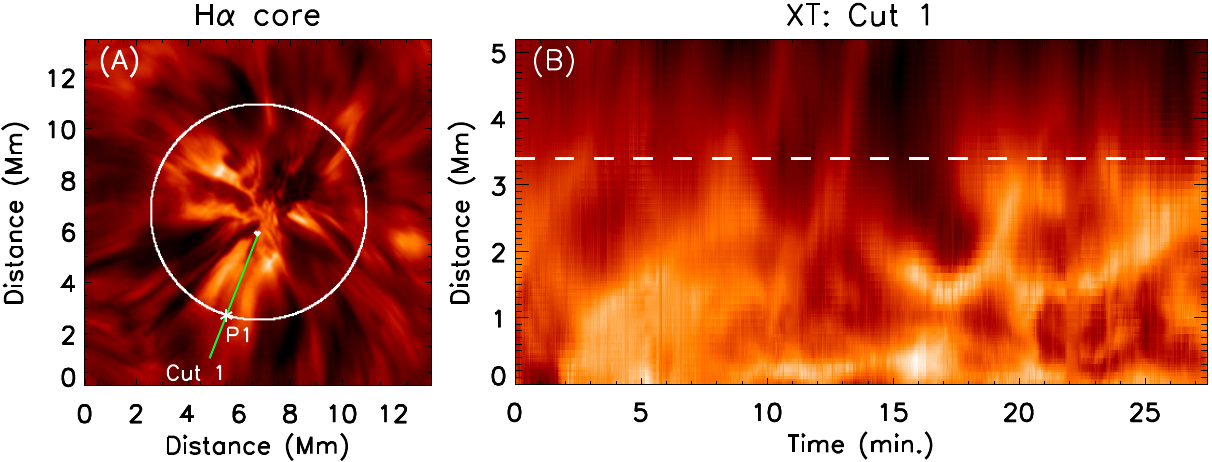}
\includegraphics[angle=90,clip,width=8.5cm]{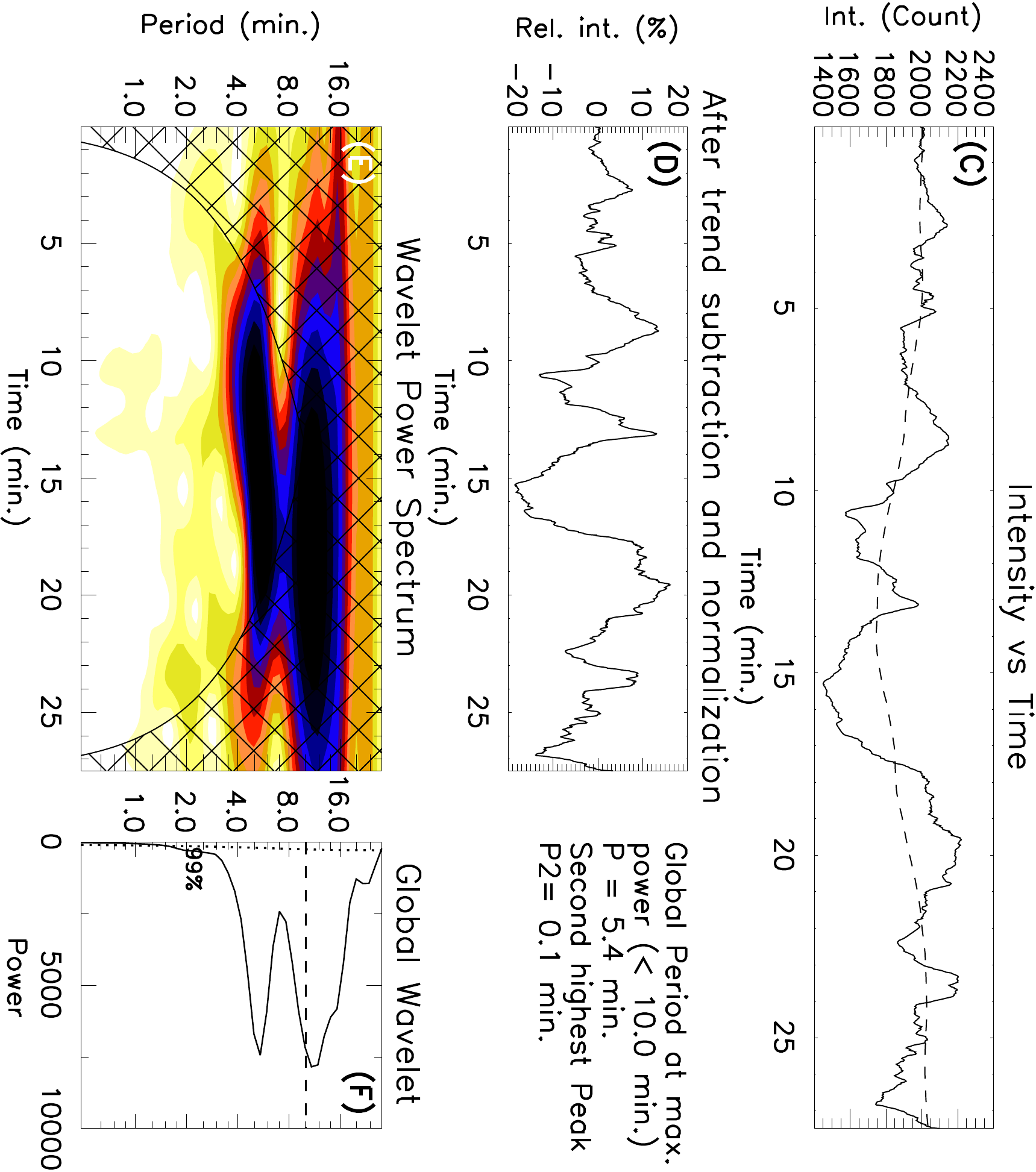}
\caption{(A) H$\alpha$ core image. Other panels are similar to those in Figure~\ref{xt_s6_cut1}.}
\label{xt_s3_cut1} 
\end{figure}

\subsection{Space-Time Plots and Wavelet Analysis}
We have generated space-time plots to study if the compressible periodic disturbances are propagating along the elongated structures in the 
network region. Artificial slits are placed radially outward from the center of the rosette structure as shown by a green solid line 
over the H$\alpha$+0.906~\r{A} image in Figure~\ref{xt_s6_cut1}A. The corresponding space-time plot is displayed in Figure~\ref{xt_s6_cut1}B  
and shows a few alternating dark ridges at the top. The propagation speeds calculated from the slope of one of the ridges is around 120 km s$^{-1}$. 
These ridges correspond to transient events like
Rapid Blueshifted Excursions (RBEs) and Rapid Redshifted Excursions (RREs) which are on-disk absorption features generally seen in the red and blue wings 
of chromospheric lines \citep{2008ApJ...679L.167L, 2009ApJ...705..272R}.
Using the same dataset, RBEs and RREs from this region have already been studied by \citet{2015ApJ...802...26K}. 
These events have the appearance of high speed jets or blobs and are generally directed outward from a magnetic network bright point 
with speeds of 50 - 150 km s$^{-1}$. They can be 
heated up to transition region (or even coronal) temperatures with a lifetime of 10 -- 120~s and are believed to 
be the on-disk counterparts of Type II spicules \citep{2014ApJ...792L..15P,2015ApJ...802...26K, 2015ApJ...799L...3R,2016arXiv160204820H}. 
We  select many locations around the network concentrations and find clear signature 
of RBEs and RREs  repeatedly appearing around the same place. Within $\sim$28 minutes of 
our observations they occur 1--15 times (intensity decreases $\gtrsim1 \sigma$) at same location with an average of 3-5 times. 
A closer inspection of the online movie shows the clear presence of such transients.

The results of wavelet analysis for the lightcurve from the row marked by a dashed line in Figure~\ref{xt_s6_cut1}B, corresponding to position P1 
marked in panel A, are shown in panels C to F of Figure~\ref{xt_s6_cut1}. Panel C displays the original light curve (solid line) and 
the background trend (dashed line), while panel D displays the relative intensity as defined in Section~\ref{power_period}. 
Panels E and F display the wavelet and global wavelet power spectra. The cross-hatched region in the wavelet plot corresponds to the 
Cone Of Influence (COI) where the periods identified are not reliable due to the finite length of the time series. The dotted line in 
the global wavelet plot corresponds to the 99~\% significance level assuming a white noise \citep{1998BAMS...79...61T}. 
The top two periods identified are also listed in the Figure. Peaks are found at 9, 4.5, and 2.5 minutes in the global wavelet power. 
Similar analysis performed over this region
in the H$\alpha$ core shows
a peak in power at 5.4~minutes (Figure~\ref{xt_s3_cut1}). 
Figures~\ref{xt_s6_cut1} and \ref{xt_s3_cut1} indicate the presence of quasi periodic fluctuations in intensity.
The fluctuations in H$\alpha$ core are probably caused by the longer lifetime of mottles (3 -- 15 minutes, \citet{2012SSRv..169..181T}).
We emphasize that we placed several slits in this region (both in the H$\alpha$ core and H$\alpha+$ 0.362 \r{A} scan positions) 
and our analysis detects oscillation periods around 3 -- 9 minutes. However, the nature of the ridges is not, generally, periodic but rather
quasi-periodic in nature. 
The dark ridges generally show high intensity drops compared to the background, which could be attributed to the outward motion of the mottles.

\begin{figure}[htbp]
\centering
\includegraphics[angle=90,trim = 5mm 42.9mm 3mm 29.4mm, clip,width=7.0cm]{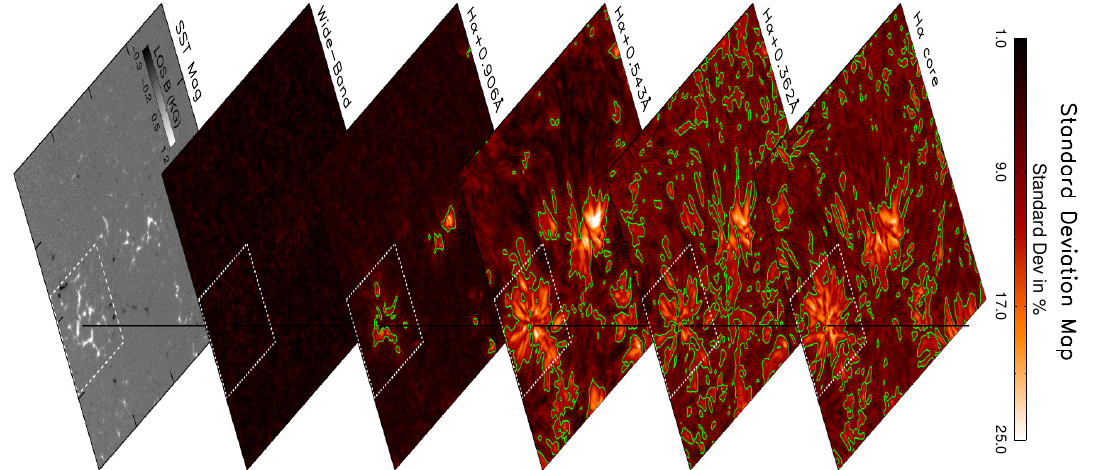}
\caption{Standard deviation maps in different layers constructed from the normalized \% standard deviation of the intensity time series at each pixel. 
The corresponding magnetogram is also shown at the bottom. The green contours enclose regions with a standard deviation of 10\% or more.}
\label{std_map} 
\end{figure}
\subsection{Standard Deviation}
We also measure the standard deviation of intensity at each pixel and construct normalized percentage standard deviation maps. 
The normalized percentage standard deviation (S) is estimated at each pixel by following S = ${I_{std}}*{I_{avg}}^{-1}*100$, 
where ${I_{std}}$ and ${I_{avg}}$ are the standard deviation and average intensity, respectively. 
The constructed maps are shown in Figure~\ref{std_map} for different layers. It is clear that close to network regions where 
we observe transients like dark mottles and RBEs, the normalized percentage standard deviation is quite high. 
A continuous periodic oscillation of 10\% amplitude (with any period) gives a standard deviation of 7\%. 
In the figure, the green contours outline the regions with normalized percentage standard deviation of 10~\% or more.

\begin{figure*}[htbp]
\centering
\includegraphics[angle=90,clip,width=8.0cm]{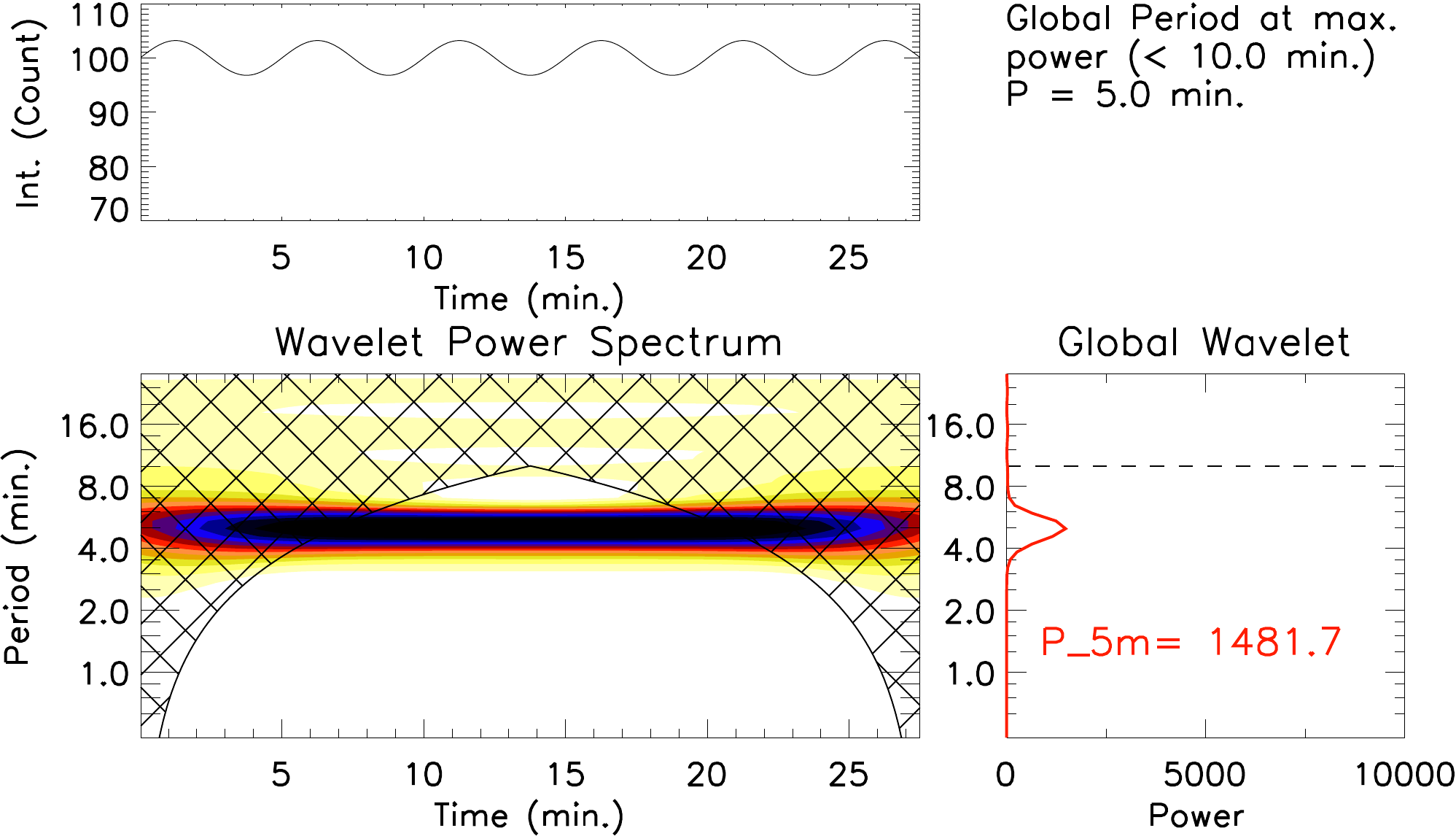}
\includegraphics[angle=90,clip,width=6cm]{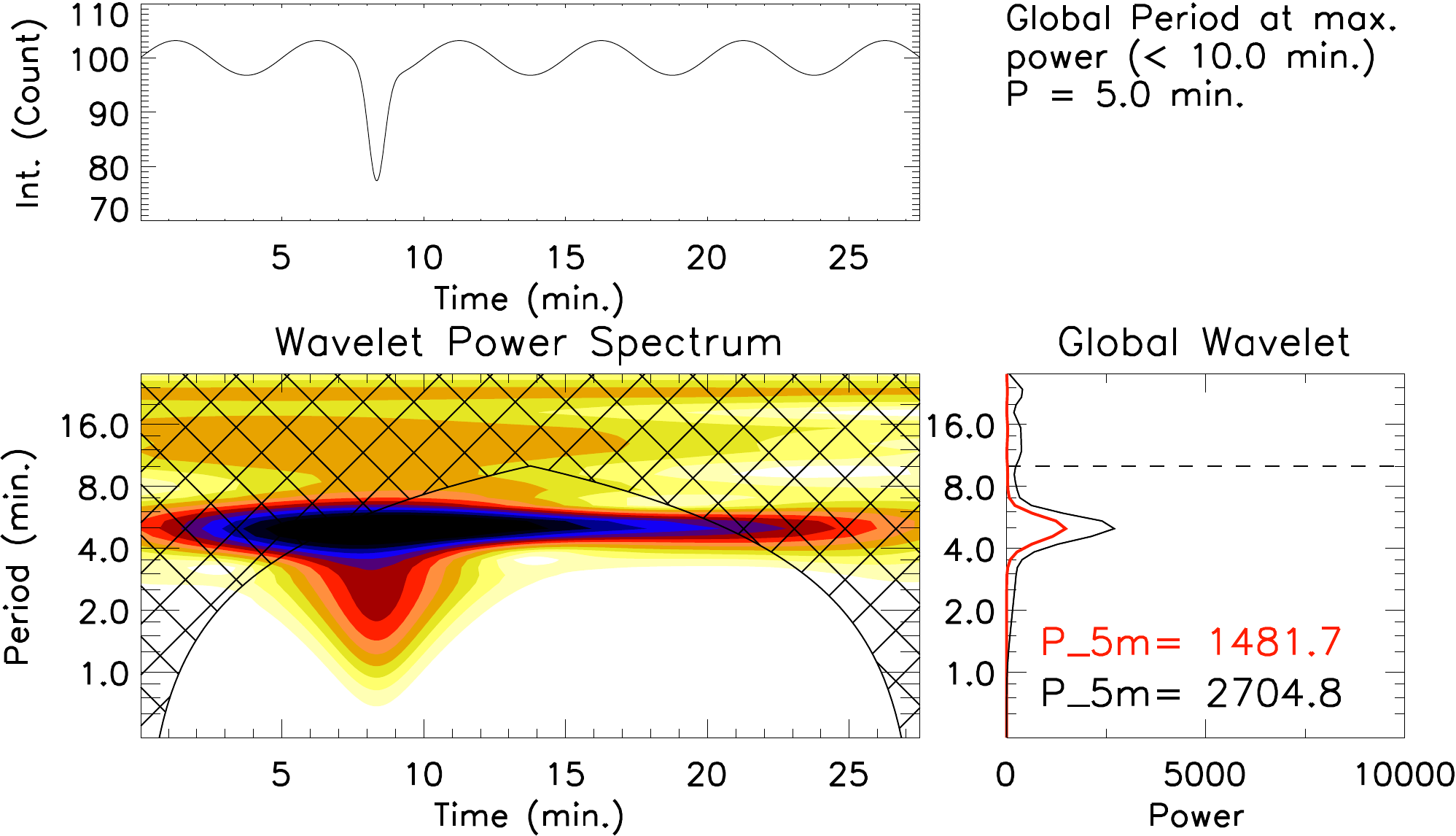}\includegraphics[angle=90,clip,width=6cm]{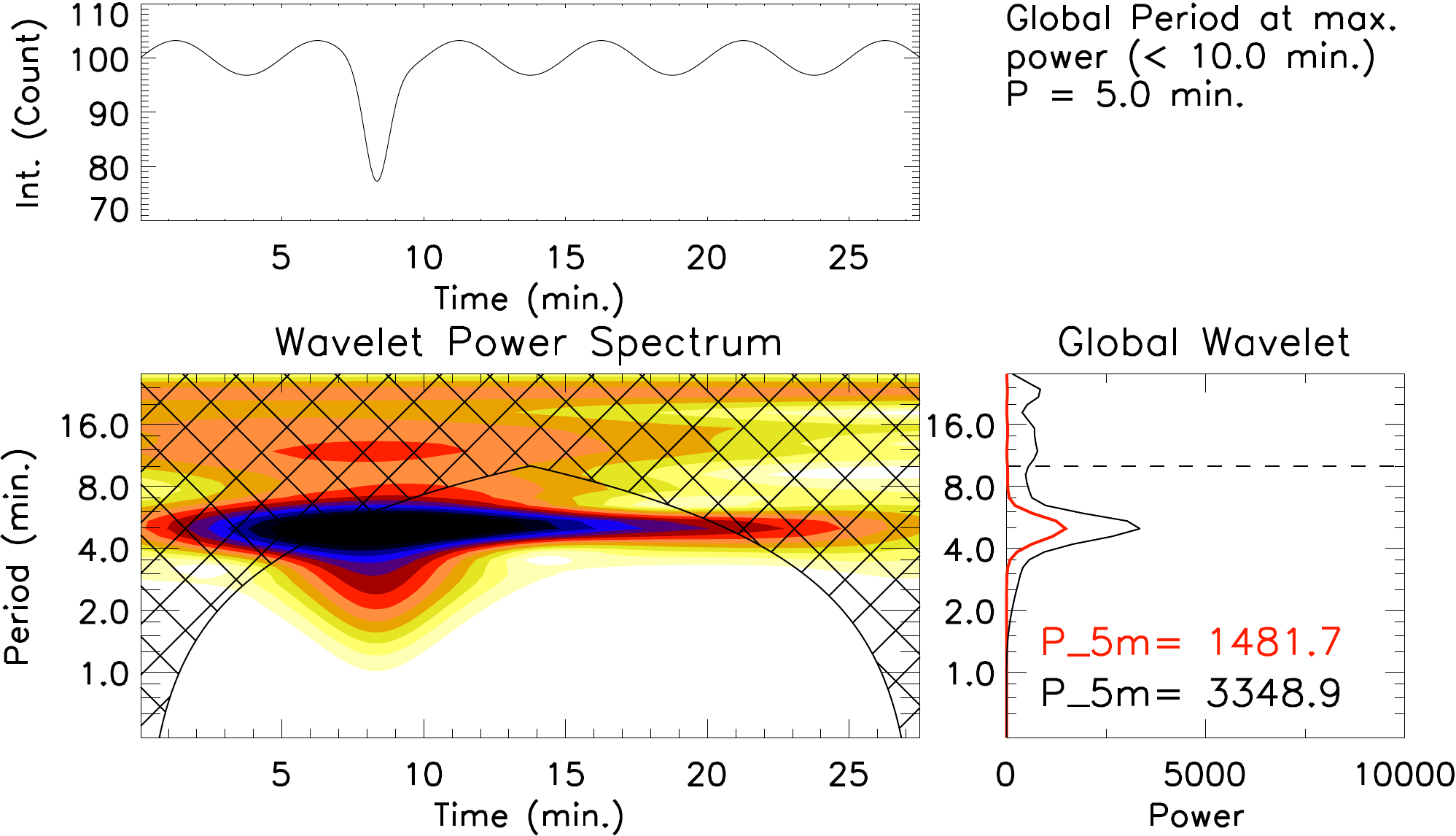}\includegraphics[angle=90,clip,width=6cm]{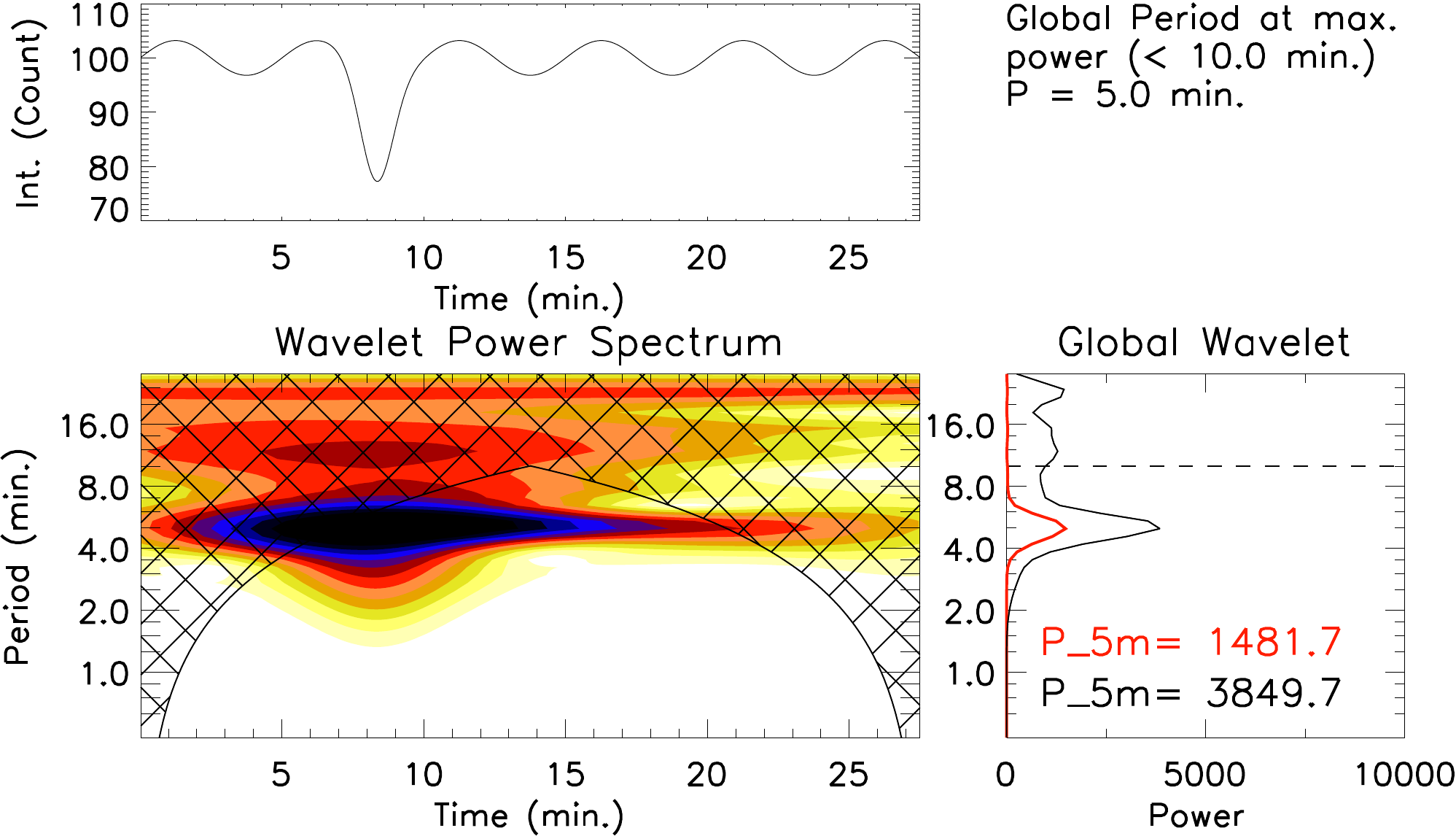}
\includegraphics[angle=90,clip,width=6.0cm]{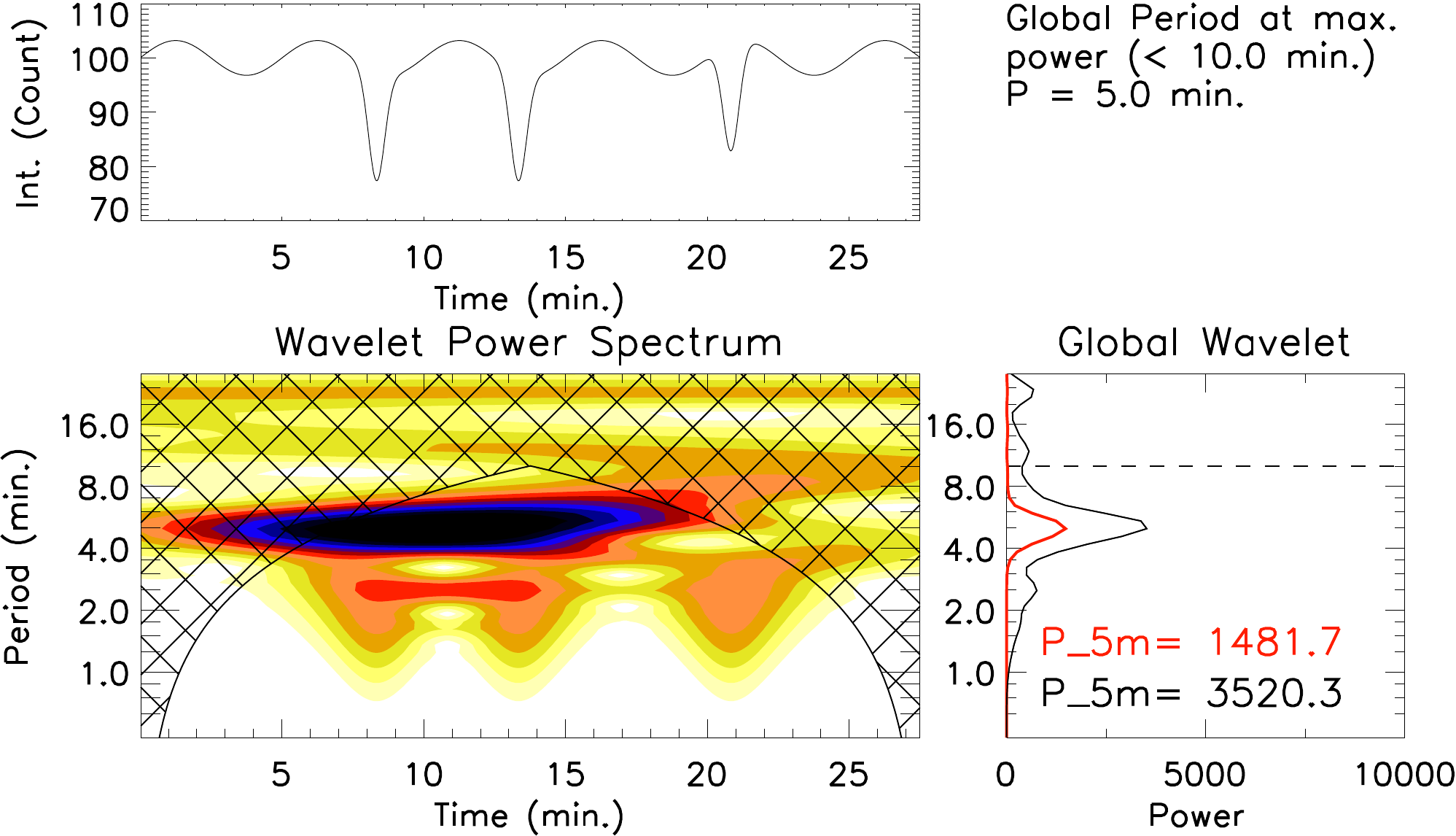}\includegraphics[angle=90,clip,width=6.0cm]{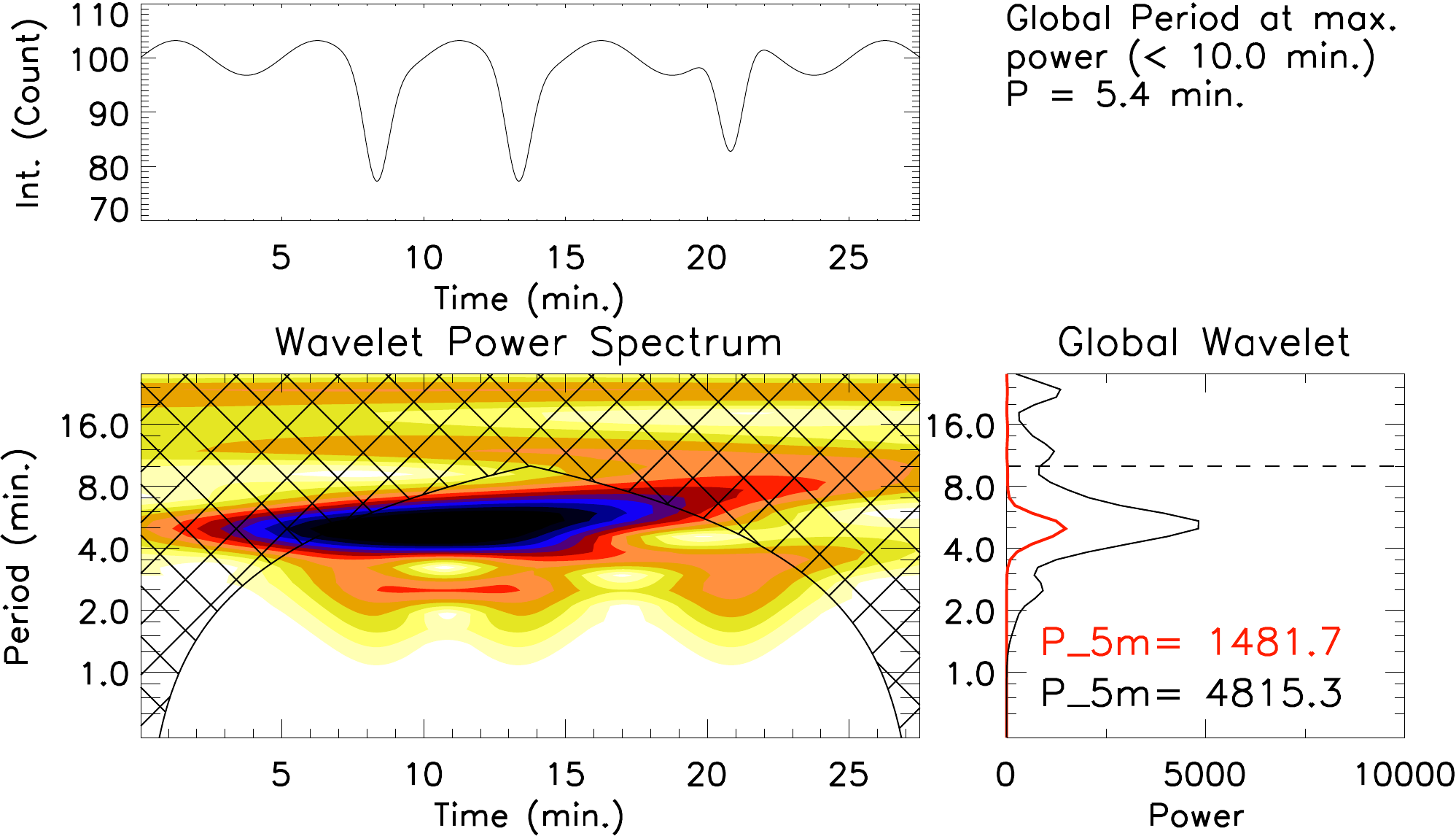}\includegraphics[angle=90,clip,width=6.0cm]{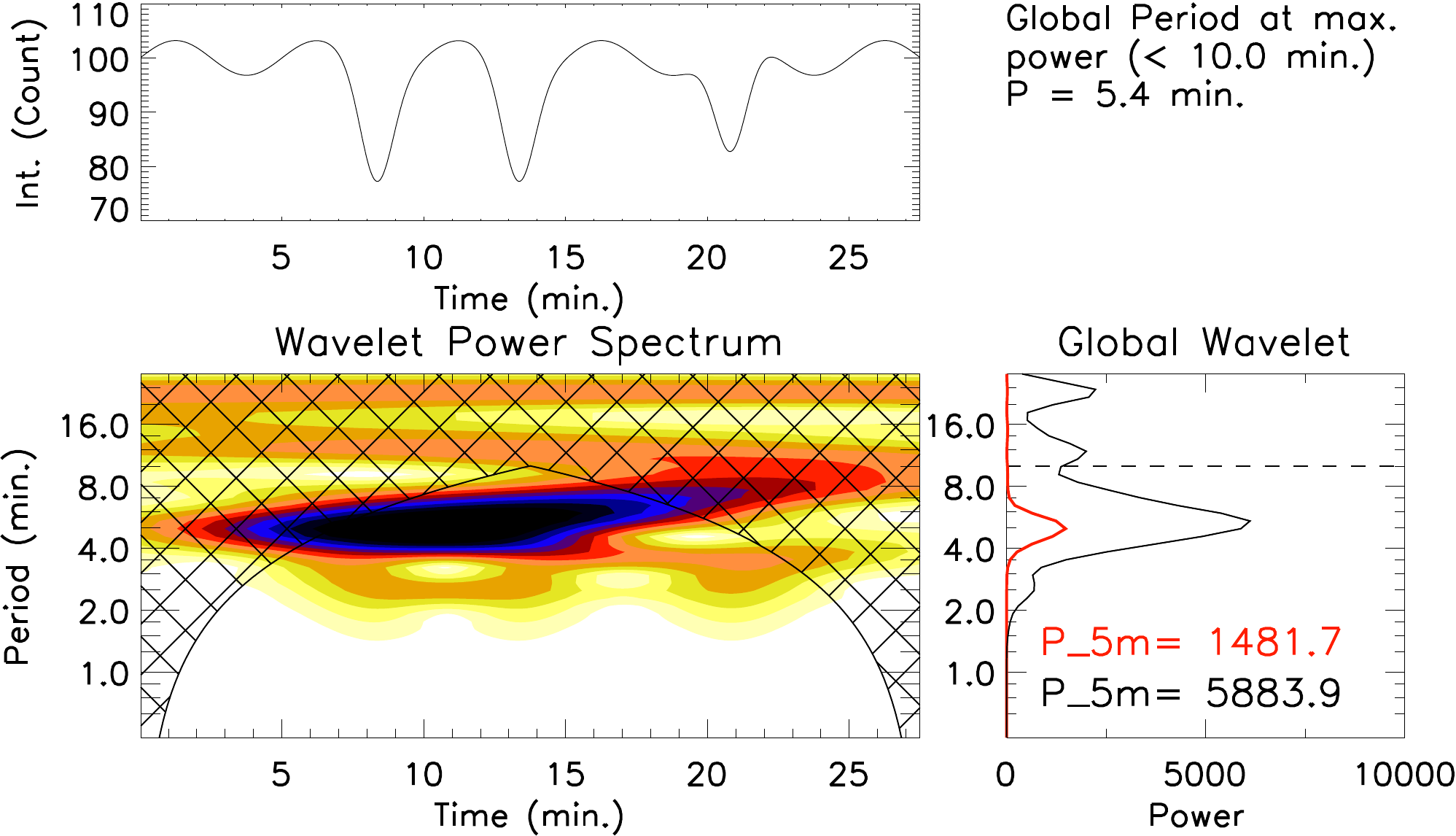}
\includegraphics[angle=90,clip,width=6.0cm]{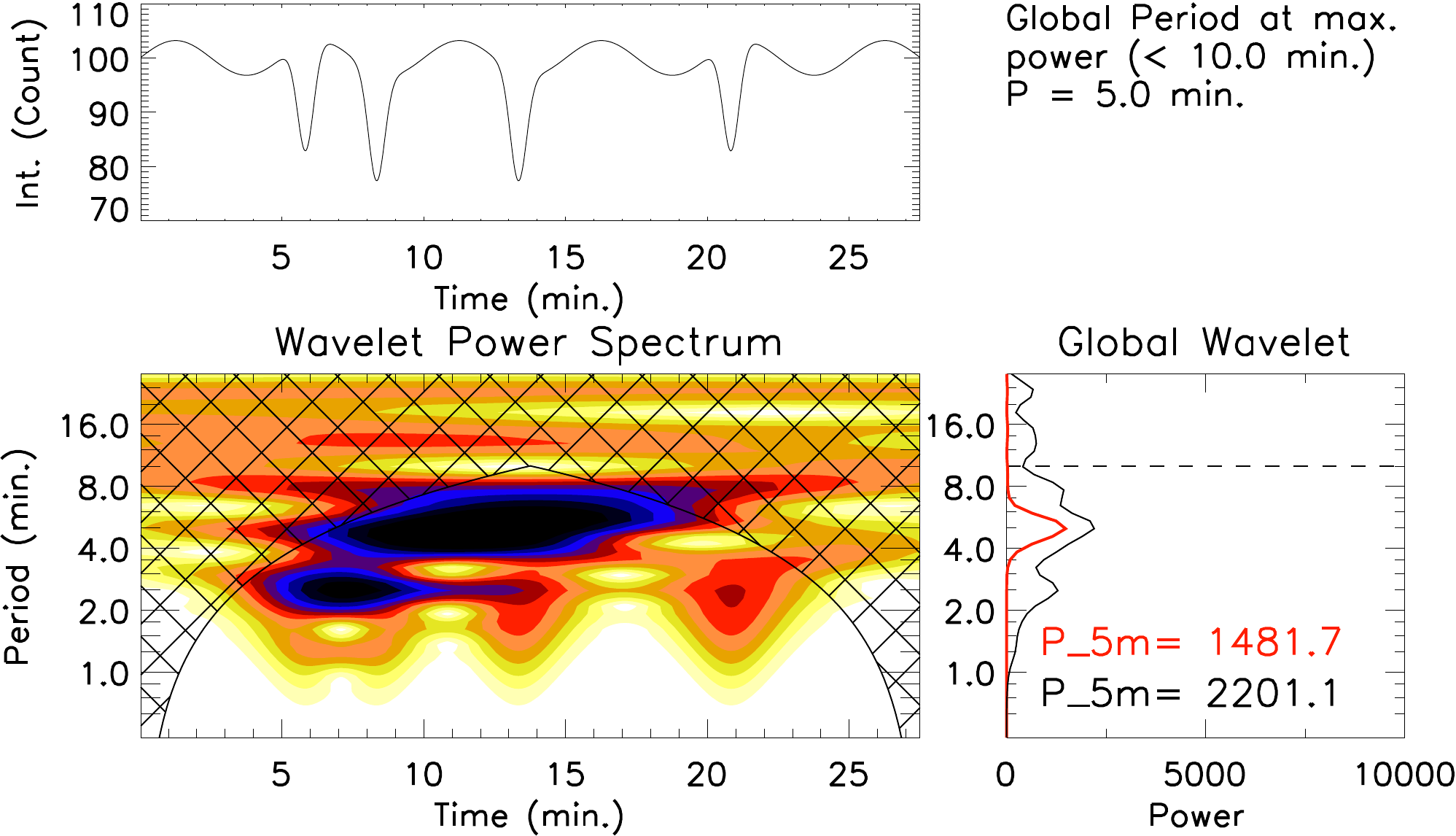}\includegraphics[angle=90,clip,width=6.0cm]{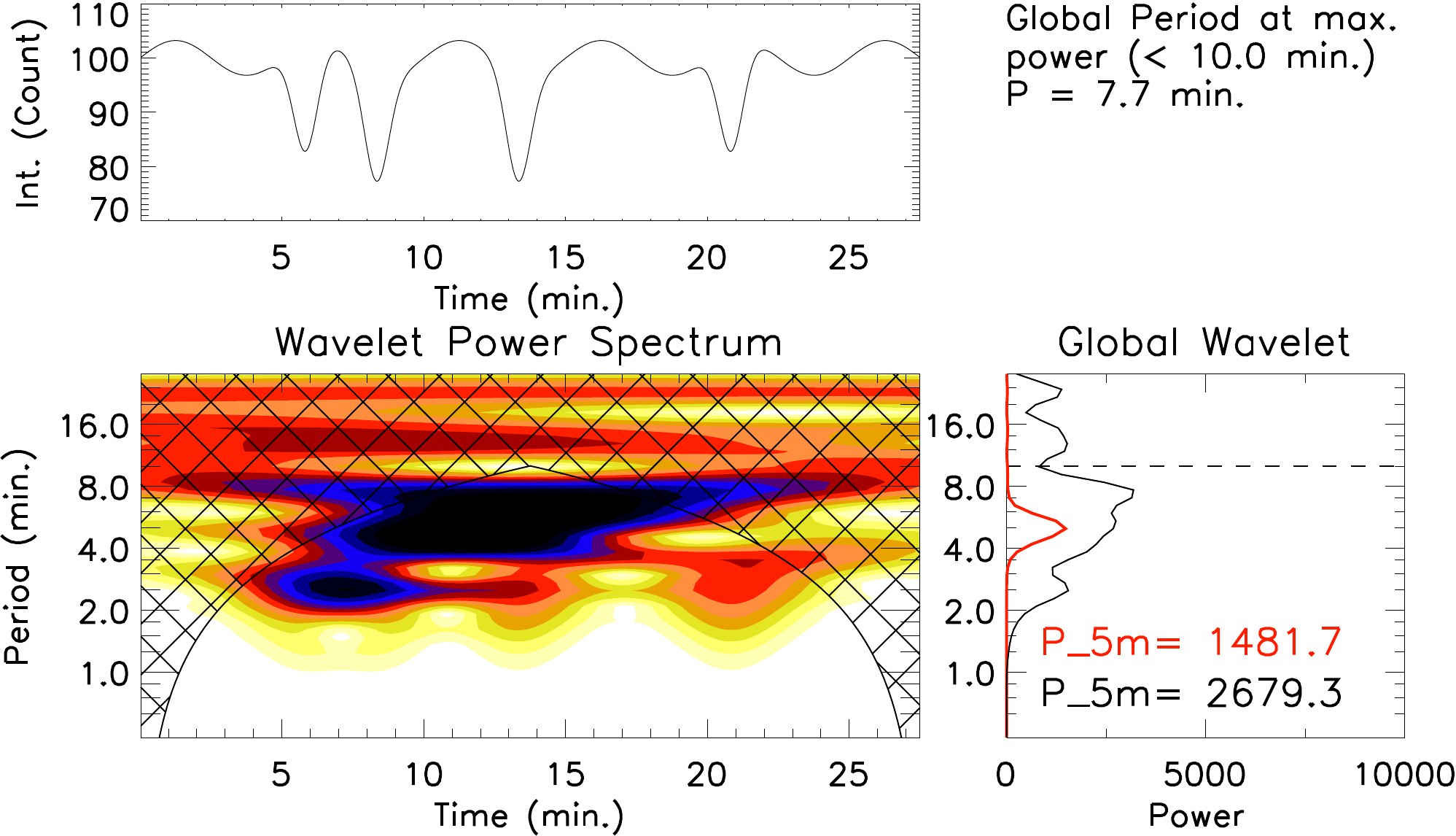}\includegraphics[angle=90,clip,width=6.0cm]{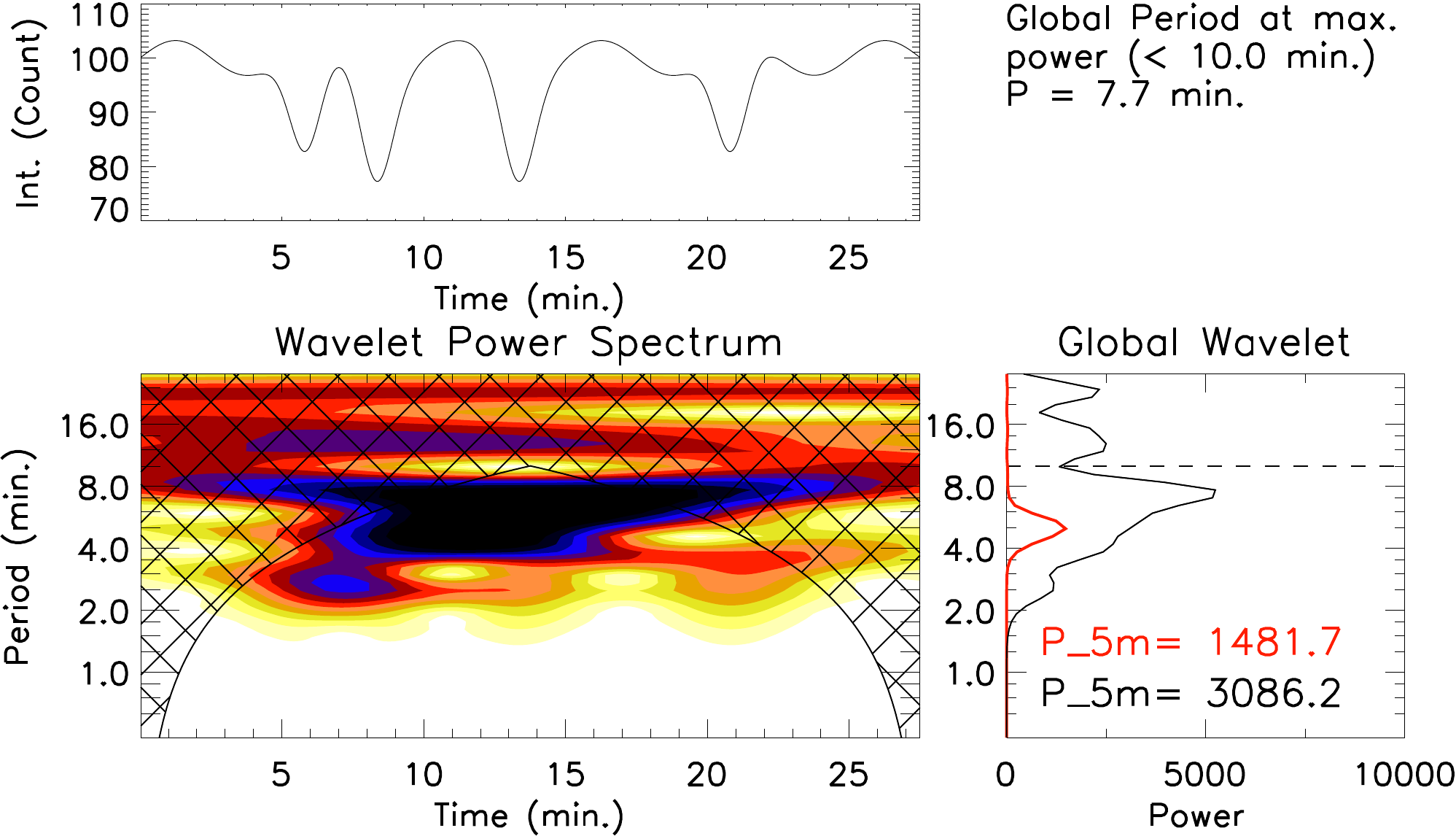}
\caption{The results of wavelet analysis for the artificially generated ligthcurves. 
Description of different panels is similar to that in Figure~\ref{xt_s6_cut1}. 
\textit{Top panel:} Wavelet analysis results for a lightcurve with a periodic sinusoidal signal of 5 minutes.
\textit{Other panels:} Wavelet analysis results for several ligthcurves artificially generated by convolving Gaussian-shaped dips in intensity with the periodic signal shown in the top panel. The convolved dips are randomly separated in time with the repetition times of 1, 3 and 4 across different rows. The FWHM of the dips has been kept at 40, 60 and 80~s across the three columns.
The amplitudes of the sinusoidal wave and the Gaussian dips are kept at 3.2 \% and 20\%, respectively, to a constant background.}
\label{wavelet_t1} 
\end{figure*}

\subsection{Artificially Generated Time Series and Wavelet Analysis}
\label{art_sig}
In this subsection we explore the signatures that will be produced in the power spectrum of a bursty signal. 
Suppose the observed variations in intensity are due to transient phenomena like RREs and mottles. We find that  RREs 
lower the intensity by 5 -- 30~\% below the background and have a lifetime of 10 to 120~s. 
Chromospheric mottles live for 3 -- 15 minutes \citep{2012SSRv..169..181T}  and cause a decrease in intensity of 10--50~\%.
Here, we have generated artificial time series to investigate the affect of non-periodic signals superimposed upon background 
oscillatory phenomena (e.g., as captured in our observation).
Our main motivation is to compare the oscillation power between the network regions (where RBEs, RREs and mottles are present) and the internetwork regions.
We have considered two different cases: the first case models the effects of transients in the power of photospheric wavelength channels, whereas the second case models the impact of transients in the power obtained from chromospheric channels.

Case-1 (photospheric channels): 
First, we have generated an artificial time series using a sinusoidal signal with a 5 minute period, which we found to be the dominant period in the photosphere (also it is well known).
We found that the averaged normalized percentage standard deviation  in the photospheric internetwork regions is $\sim$2.25~\% (see Figure~\ref{std_map}). 
In order to match with the observed normalized percentage standard deviation, we have selected the amplitude of the sinusoid periodic signal to be 3.2~\% with respect to a constant background (ignoring all the noise and other high- and low-frequency fluctuations). 
We then performed wavelet analysis on this artificial signal to compute the global wavelet power spectrum as a reference which is shown in the top panel in the Figure~\ref{wavelet_t1}. 
This was followed by introducing random fluctuations in the same 5 minute
periodic signal.
The repetition and the amplitudes of the random fluctuations were selected such that they can mimic the observed lightcurves (an example of the observed lightcurve is shown in Figure~\ref{xt_s6_cut1}). 
We find that many of the RREs show intensity drops between 10 to 25\% compared 
to the background intensity whereas some of the weak and strong RREs have intensity 
drops less than 5\% and more than 30\% respectively. They generally occur repeatedly at the same location (close to the network) with an average of 3--5 times in $\sim$28 minutes.
Keeping in mind the observed distribution of the transients (RREs), we have produced lightcurves while introducing sudden fluctuations (Gaussian-shaped dips) with random repetitions (1--5 times) in the same 5 minute
periodic signal.
We generated forty five lightcurves while changing the amplitudes (10\%, 20\% and 30\%) and temporal width (FWHM of 20, 40, 60 and 80~s) of the Gaussian-shaped dips. We then subjected these modified lightcurves to wavelet analysis for computing the power spectra.
Some representative examples (only for the Gaussian dips of intensity amplitudes drops of 20\%  with 1, 3 and 4 time repetitions and FWHM of 40, 60 and 80~s)  are shown in the Figure~\ref{wavelet_t1}. 
We compare the power of 5 minutes oscillation of the reference periodic-signal (P\_5m in red) with the power of the same signal with fluctuations (P\_5m in black). 
Our analysis shows that the power of the 5 minutes period is enhanced 1.1--6.8 times due to the presence of 
RBE-like random fluctuations in intensity and lifetime. 
The enhancement in power is dependent on the amplitude, temporal width, repetition and also the temporal location (phase) of Gaussian dips. 
We should point out that we have compared the observed 3-minute power between the regions of enhanced power (network) 
and internetwork regions  in  H${\alpha}$ + 0.906 \r{A} and find the the enhancements in power is around 2--5 times compared to the internetwork regions. 
Additionally, we find that the  power in the period-band of 2--9 minutes 
is enhanced which is similar to the observed power-distribution.

\begin{figure*}[htbp]
\centering
\includegraphics[angle=90,clip,width=8.0cm]{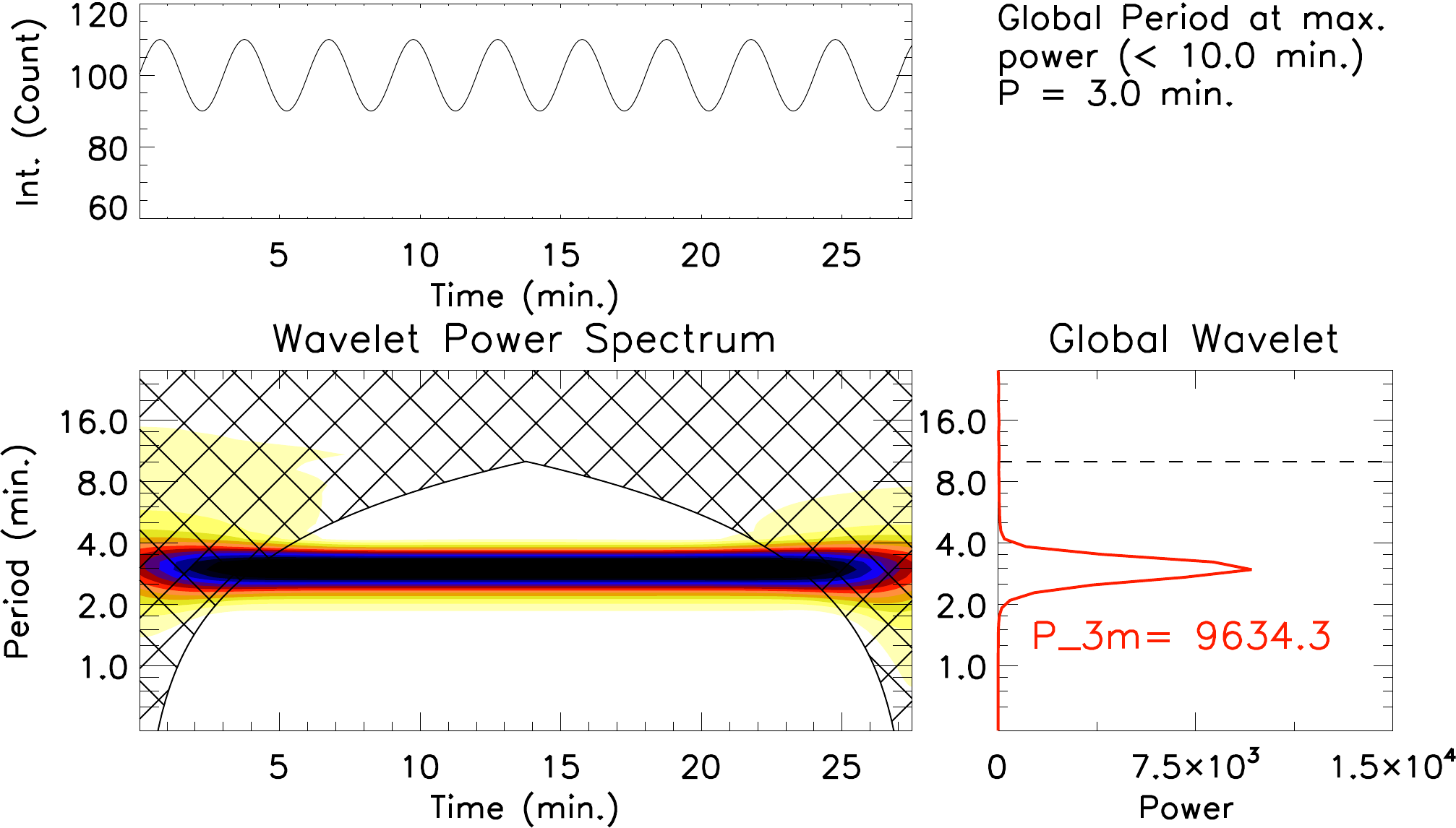}
\includegraphics[angle=90,clip,width=6.0cm]{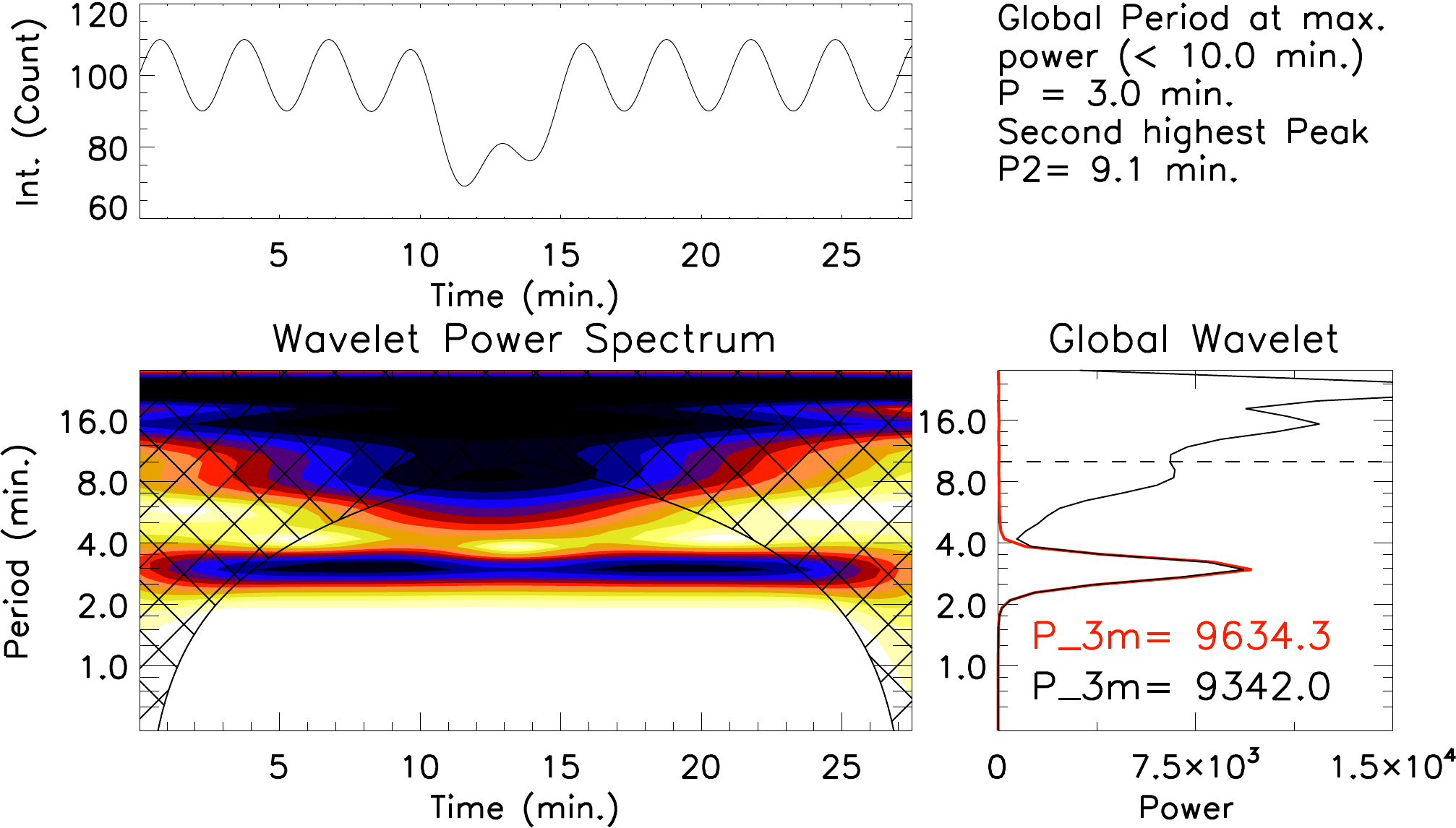}\includegraphics[angle=90,clip,width=6.0cm]{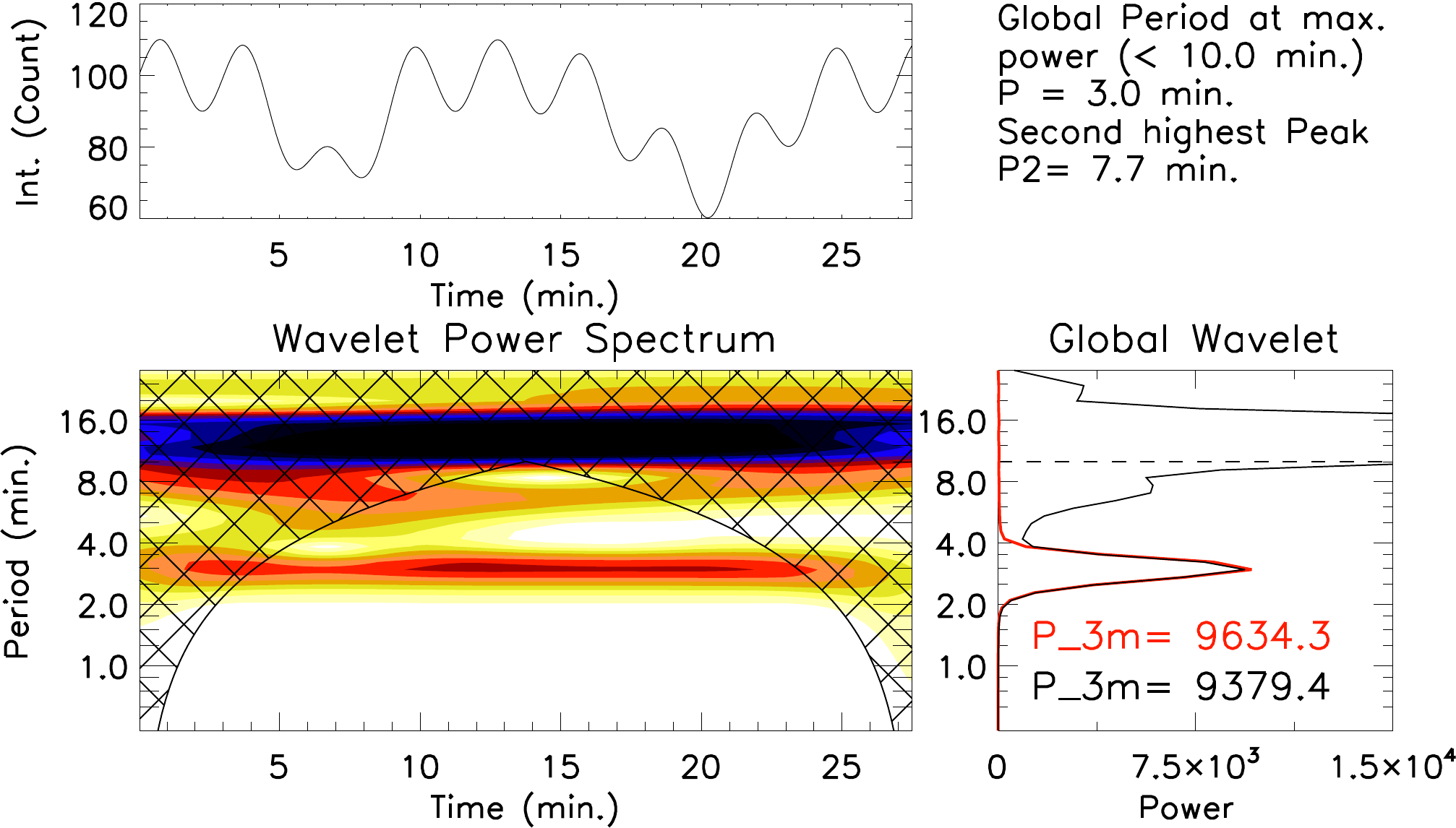}\includegraphics[angle=90,clip,width=6.0cm]{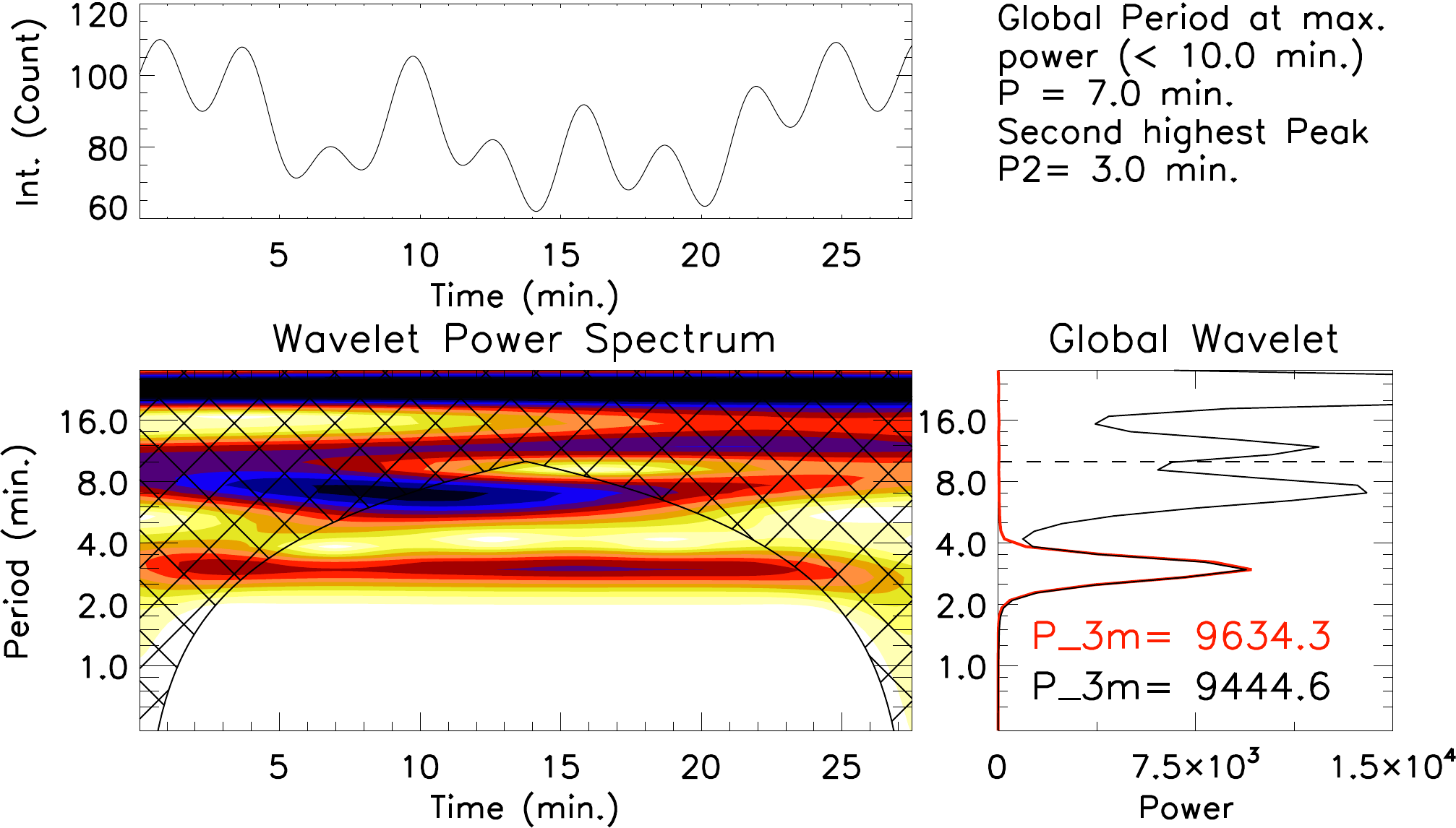}
\includegraphics[angle=90,clip,width=6.0cm]{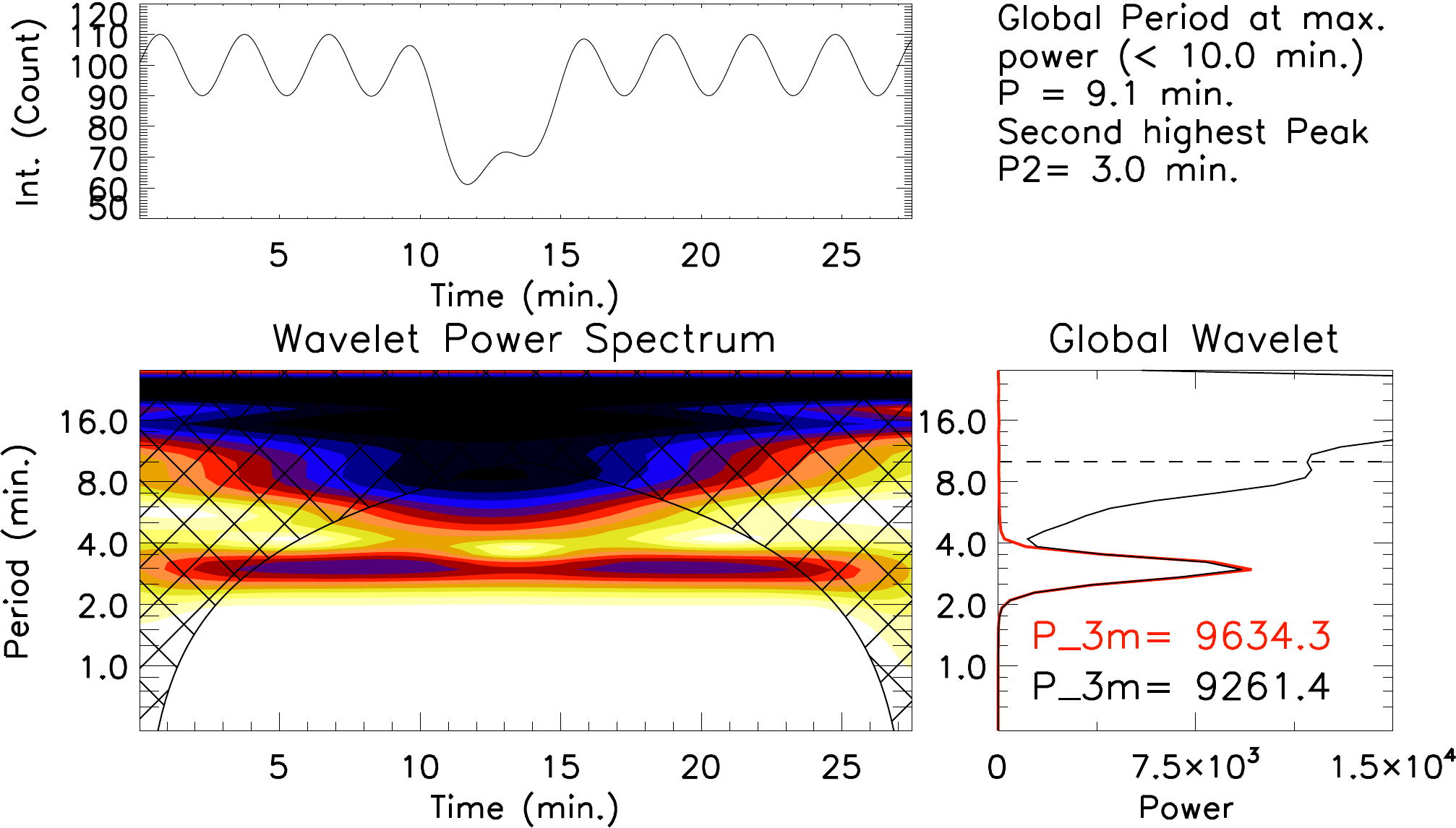}\includegraphics[angle=90,clip,width=6.0cm]{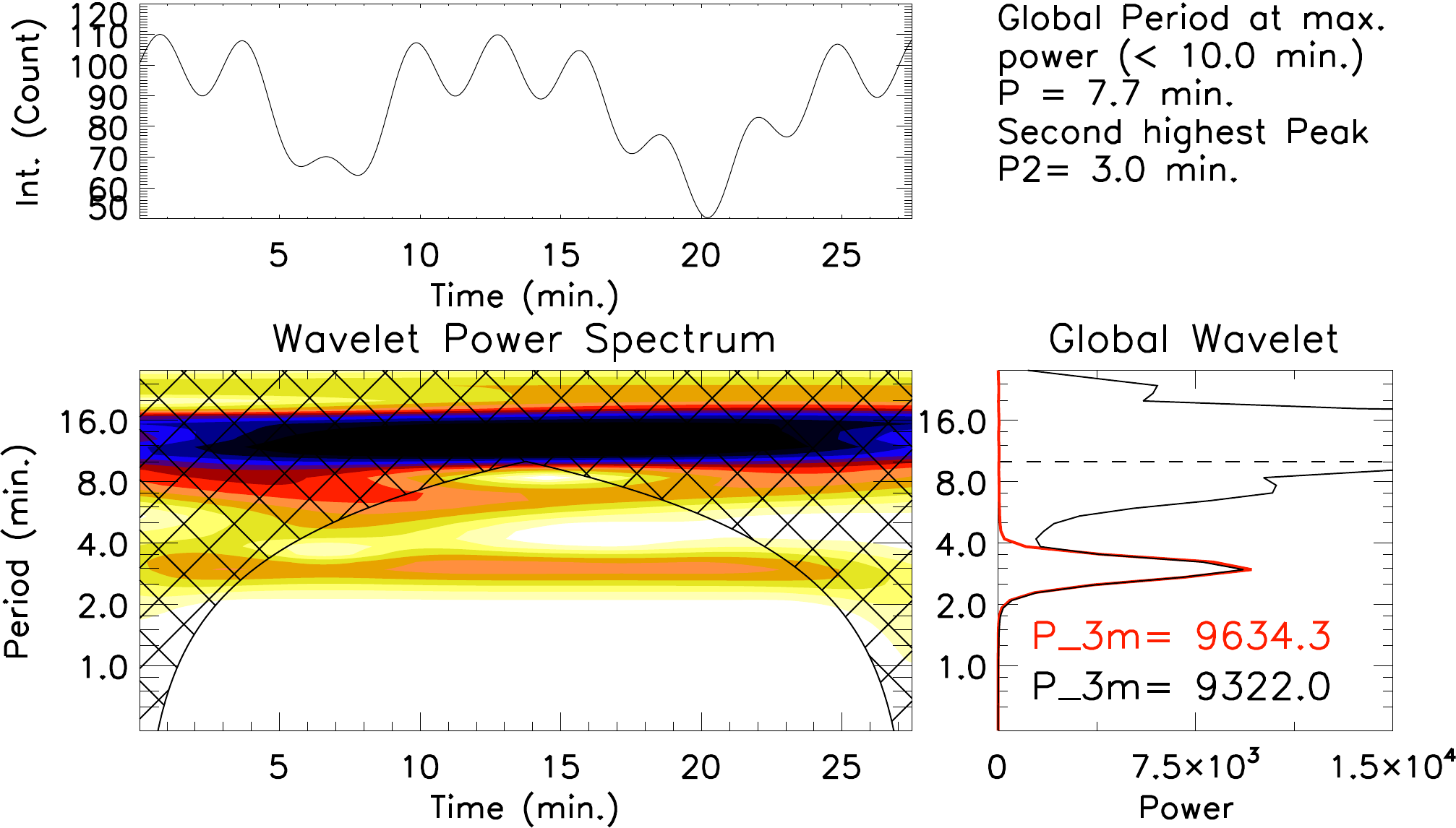}\includegraphics[angle=90,clip,width=6.0cm]{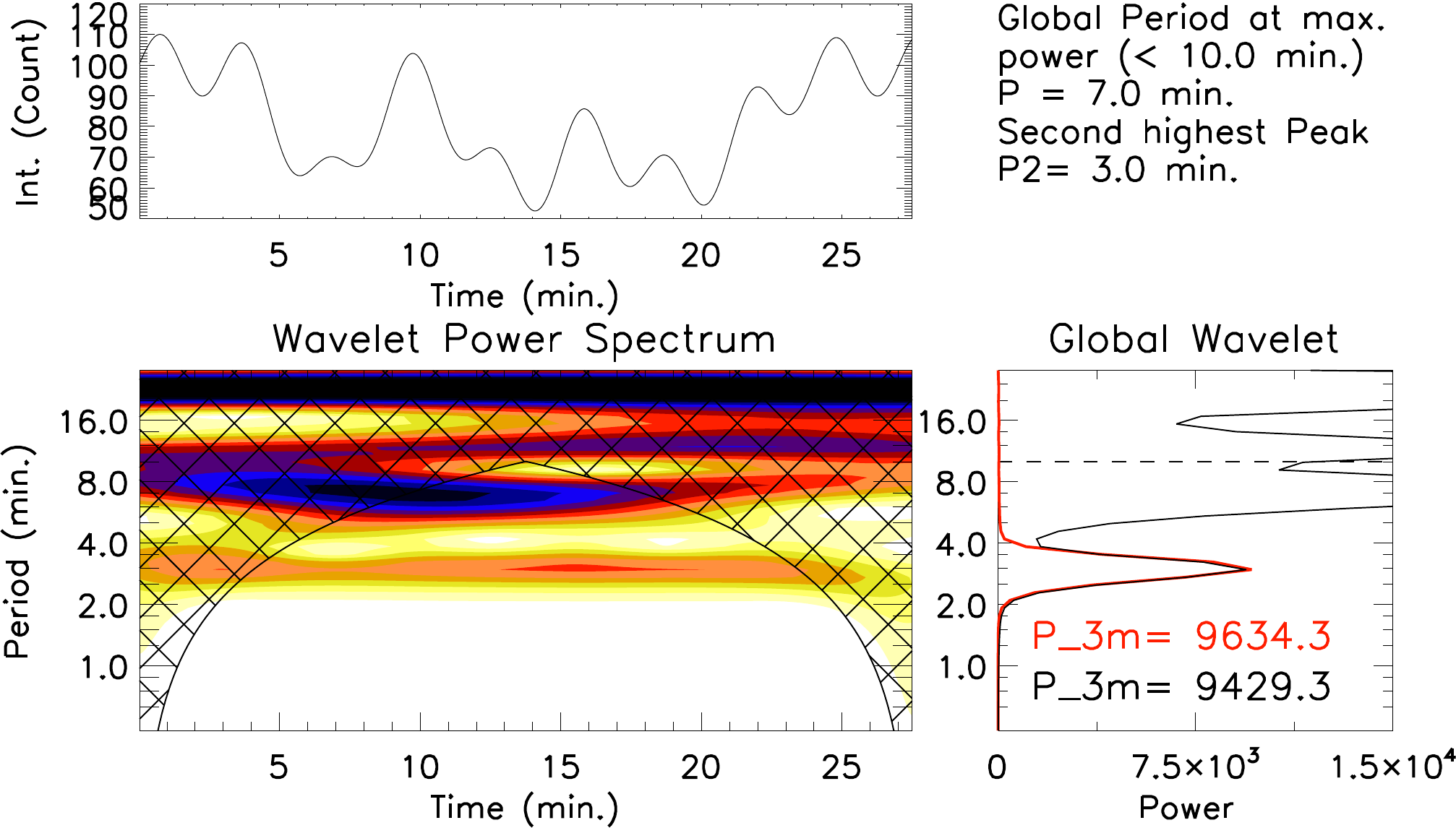}
\caption{The results of wavelet analysis for the artificially generated ligthcurves. Description of different panels is similar to that in Figure~\ref{xt_s6_cut1}. 
\textit{Top panel:} Wavelet analysis results for a lightcurve with a periodic sinusoidal signal of period 3 minutes.
\textit{Middle panels:} Wavelet analysis results for several artificially generated lightcurves made by convolving Gaussian-shaped dips of FWHM 3 minutes; 3 and 5 minutes; 3, 5 and 7 minutes, with the periodic signal shown in the top panel.
The amplitudes of the sinusoidal signal and the Gaussian-shaped dips are kept at 10~\% and 30~\% with respect to the background, respectively.
\textit{Bottom panels:} Same as the middle panels but for a 40~\% amplitude of the Gaussian-shaped dips with respect to the background.}
\label{wavelet_t2} 
\end{figure*}

Case-2 (chromospheric channels): Similarly to the first case, we have generated an artificial timeseries with a periodic sinusoidal signal of period 3 minutes.
Here we selected the period of the oscillation to be 3 minutes  as the chromospheric internetwork regions (H$\alpha+$ 0.362 \r{A}) are dominated by a 3 minute period.
As before, to compare with the observations, we have selected the amplitude of the sinusoid to be 10\% (we find the average normalized percentage standard deviation is $\sim$ 7.4\% in the internetwork regions of  H$\alpha$ + 0.362 \r{A} layer) with respect to the background.
We find that the intensity drops in H$\alpha$ + 0.362~\r{A} due to presence of mottles is around 10--50\%.
We have produced twenty seven lightcurves while introducing random fluctuations (1--3 Gaussian dips distributed along the whole time-series) by changing the amplitudes (20, 30 and 40\%) and temporal width (FWHM of 3, 5 and 7 minutes) followed by wavelet analysis 
to compute the power. 
Few examples (for the Gaussian dips with amplitude of 30 and 40\% only) are shown in Figure~\ref{wavelet_t2}. We compare the power of 3 minutes oscillation of the pure periodic signal (P\_3m in red) with the power of the same signal with fluctuations (P\_3m in black). 
Our analysis shows that the power of the 3 minute period gets suppressed  2--6\% (though the observed magnetic shadow region show around 60--70\% decrease of the power of 3 minutes oscillation compared to internetwork regions) due to the presence of 
random fluctuations like mottles. We also noticed that the power in the period band of 5--9 minutes is generally enhanced due to this kind of sudden fluctuations. 
Hence, a sudden drop in intensity with a random distribution in time can lead to significant power at different periods. 
One important thing to note here is that the periods mainly depend on the distribution of the intensity drops and they are generally 
longer than their FWHM. 
We should point out that sometimes the power gets enhanced depending on the phase of the Gaussian dips with respect to the continuous 3 minute periodic sinusoid.

\begin{figure}
\centering
\includegraphics[angle=90,clip,width=8.5cm]{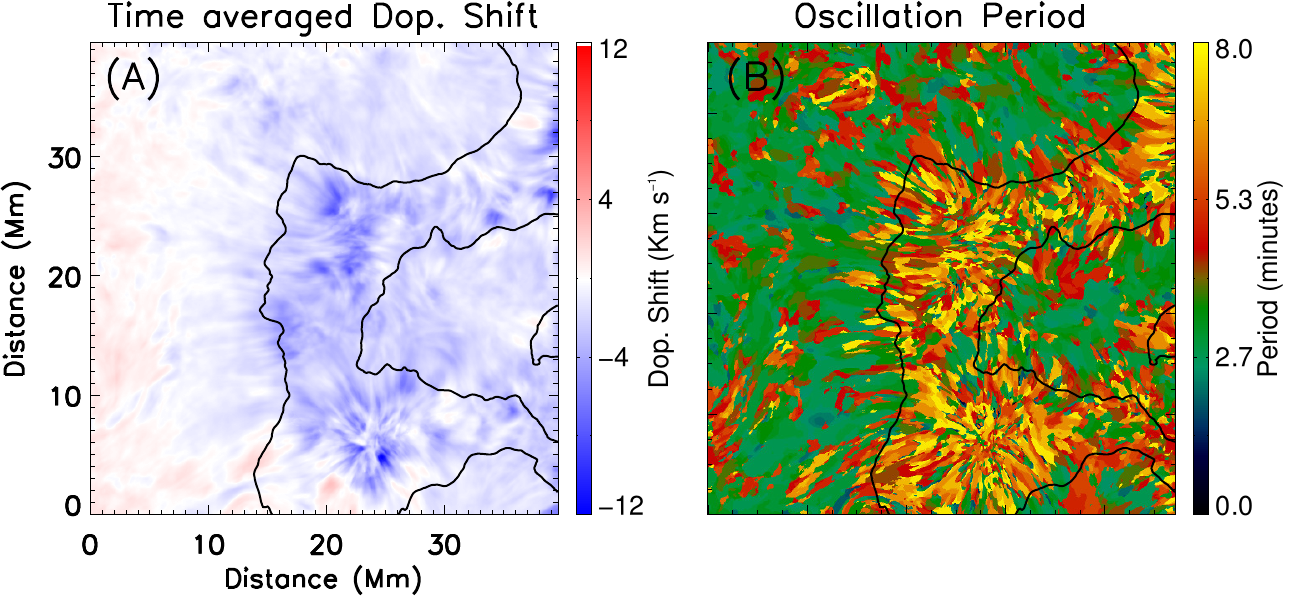}
\caption{(A): Time-averaged Doppler velocity map of H$\alpha$ line. (B): Distribution of dominant periods in Doppler velocity oscillations. 
Contours on both plots represent a dominant period level of 4.5 minutes. The contours are calculated after smoothing the image in panel B.}
\label{avg_doppler} 
\end{figure}
\begin{figure*}
\centering
\includegraphics[angle=90, clip,width=3.04cm]{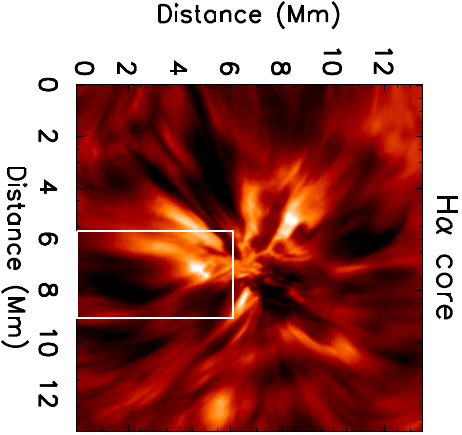}\includegraphics[angle=90,trim = -10.70mm 0mm 0mm 0mm, clip,width=14.7cm]{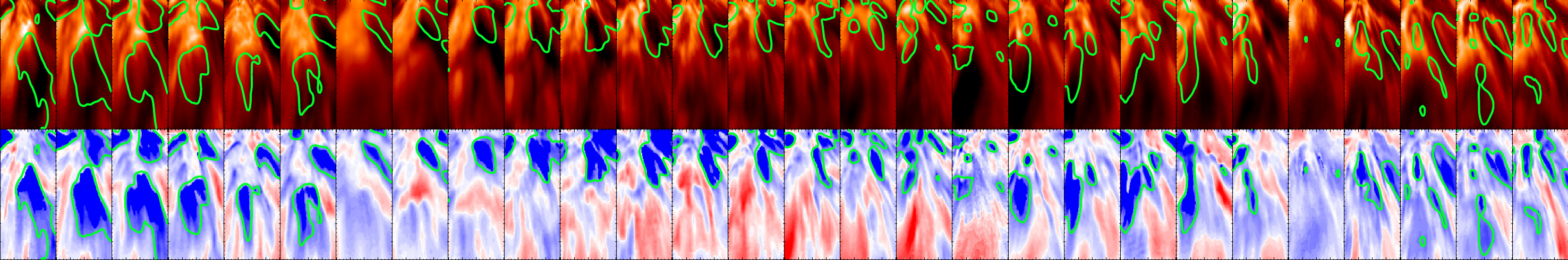}
\caption{In the extreme left panel, the white rectangular box marks our region of interest.  Right panels show time evolution of the portion inside the white rectangular box covering a few dark mottles. 
The top and bottom rows display the intensity and Doppler velocity in H$\alpha$ core. Each frame is separated by a 1-minute interval. Green contours on intensity correspond to -6~km~s$^{-1}$ Doppler velocity.}
\label{core_dop} 
\end{figure*}

\subsection{Time-Averaged Doppler Shift and Material Outflows.}
The time-averaged Doppler velocity provides very important information on the statistical properties of the dynamics. Figure~\ref{avg_doppler}A shows the 
time-averaged Doppler velocity map of the whole FOV. The overplotted contours outline a region with a dominant periodicity of 4.5~minutes as 
shown in the period-distribution map (Figure~\ref{avg_doppler}B). It can be seen that within this region, above the network, the average Doppler velocity is blue shifted ($\sim$ 5 km~s$^{-1}$). 

The evolution of a portion of the network region is shown in Figure~\ref{core_dop}. The white box marked in the left panel is our region of interest for temporal variations. 
The upper panels display the intensity and the bottom panels display the Doppler velocity as captured in one-minute intervals. 
This figure shows that when dark mottles first start appearing, they are blueshifted but with time the mottles evolve and become bright and redshifted. 
It is possible that mottles are nothing but strong material outflows like Type I spicules. They appear similar to Type I spicular flows following parabolic paths. 
The material moves outwards causing a blueshift which turns to redshift when the material falls back on the solar surface.

\section{Discussion}

As pointed out in the introduction,  the interaction between acoustic waves and the magnetic field are responsible  
to the formation of magnetic shadows and power halos \citep{2001ApJ...554..424J,2003A&A...405..769M,2007A&A...471..961M,2007A&A...461L...1V,2010A&A...524A..12K,2014A&A...567A..62K}. 
It is to be noted that using Dutch Open Telescope  H$\alpha$ observations with a cadence of 30~s, \citet{2010A&A...510A..41K,2010A&A...524A..12K} 
pointed out that there is a strong possibility that power at longer time periods ($\sim$7 minutes)
may be enhanced as a result of the lifetimes of the mottles.
Furthermore, \citet{2010A&A...510A..41K}, 
also highlighted that the observed power enhancements, at both photospheric and chromospheric heights,
may be closely related to the temporal dynamics of such transients and their lifetimes.
In this paper we have explored if transients can influence the power distribution at different heights. 
The high-cadence (1.34s) observations presented here allow us to identify and study the dynamics of transient phenomena in greater detail, 
which was previously not possible due to lower cadence ($\sim$30~s).

The quiet chromosphere is generally dominated by numerous elongated dark structures seen in H$\alpha$. 
These include rapidly changing hair-like structures known as mottles and extreme Doppler-shifted events such as RBEs and RREs 
(for details see review of \citet{2012RSPTA.370.3129R,2012SSRv..169..181T}). 
Figure~\ref{observation} and movie~1 (available online)  
reveal that these structures are associated with the regions of network magnetic-fields which appear at 
the edges of granular cells \citep{2009LRSP....6....2N}. 
It is now generally believed that the dark mottles are the disk counterparts of Type I spicules \citep{1994SSRv...70...65T,1994A&A...290..285T,1997A&A...324.1183T,2001SoPh..199...61C} 
and the RBEs are the  disk counterparts of Type II spicules \citep{2008ApJ...679L.167L,2011Sci...331...55D,2012ApJ...759...18P,2015ApJ...802...26K}.
The mottles seen in the H$\alpha$ line have mean velocities of the order of 20-40 km s$^{-1}$ and lifetimes of 3-15 minutes \citep{2012SSRv..169..181T}. 
On the other hand, the transients like RBEs and RREs generally exhibit upward motion and rapidly fade away without any signature of downward motion. 
They have shorter lifetimes (10-120 s), high apparent velocities (50-150 km s$^{-1}$), and smaller widths (150 and 700 km) \citep{2015ApJ...802...26K}.

\citet{2003A&A...402..361T} found that mottles arise at the network boundaries as bursts of material and propagate upward with a velocity around 25 km s$^{-1}$. \
They also show a tendency to occur several times at the same place with a typical duration of around 5 minutes. 
Our analysis also indicates that the mottles are jet like features originating in the network region that propagate upward. 
Figure~\ref{core_dop} shows that the footpoints of mottles display strong blue-ward shift when they originate 
but with time they fade away and small red-shifts are observed that likely correspond to material falling back along the magnetic-canopy structures. 
The average blueshift above network (see Figure~\ref{avg_doppler}) indicates that material outflows are present in that region. 
These outflows are not as strong as they are in individual time frames (see Figure~\ref{doppler}), suggesting that outflows are not continuous but rather quasi-periodic in nature. 
The normalized percentage standard deviation in the 
photosphere, where 5 minute p-modes dominate, is low (around 2.25~\%) but, above the network regions where the RREs and mottles are seen, 
is quite high (above 10\%, see Figure~\ref{std_map}). Higher values of normalized percentage standard deviation can not be explained solely by 
the presence of linear MHD waves (observations show that the slow waves generally have an amplitude of less than 5 \% \citep{2011SSRv..158..397W}). 
Numerical models show that the dark mottles observed in H$\alpha$ are due to material density enhancement \citep{2006A&A...449.1209L,2012ApJ...749..136L}. So, the fluctuations caused by the rise and fall of material in the form of transients  may be responsible for the observed high standard deviation. 

Power halos (across all period bands) manifesting in the predominantly photospheric bandpass (H$\alpha+$0.906~\AA) 
can be explained due to the occurrence of Doppler-shifted transients like RBEs and RREs.
The online movies clearly show the presence of these transients, 
particularly in the neighbourhood of the network field concentrations.
Although not strictly periodic, they occur repeatedly (3 -- 15 times in
28 minutes) at the same location and have lifetime of 10 -- 120 s.
Hence, the lifetime and distribution of Doppler-shifted RBEs (see Figure~4 of \citet{2013ApJ...764..164S}) 
can produce sufficient power enhancement in different periods as shown from artificial lightcurves 
as demonstrated in Figure~\ref{wavelet_t1} that corresponds to ``case-1'' from Subsection~\ref{art_sig} 


Imaging data from a passband centered at 0.7 \AA\ from the H$\alpha$ line core was used to produce the power-maps where quiet-Sun 
``power halos'' were positively identified \citep{2010A&A...510A..41K,2010A&A...524A..12K,2014A&A...567A..62K}. 
 This is closer to the H$\alpha$ line core when compared to the predominantly photospheric bandpass at +0.9~\AA.
Thus, we believe  that these previous power-halo detections were more affected by Doppler-shifted transients than our observations and simulations, firmly setting Doppler-shifted transients as the source of the observed halos in the 
quiet Sun.
Note that in our wide-band power maps we do not find significant power enhancement at the regions of halos as observed in the H$\alpha$ + 0.906~\r{A}. This confirms earlier report 
by \citet{2007A&A...461L...1V} who also do not find signature of power enhancements in the photospheric broadband continuum band (centered at 710 nm) close to network regions.
There is no reason that power-halos should not be observable in wide-band data as they are photospheric. There should be no difference between narrow-band observations at purely photospheric wavelengths and wide-band observations with respect to wave detection. The effective difference we find between the two is impact of the Doppler-shifted chromospheric transients.

Similarly, we believe that power from random transients could affect  the light curves and influence the power distribution.  More generally, presence of transients can leave a two-dimensional signature visible in power-maps obtained in a similar fashion. One such example is ``network aureoles'', a structure similar to the power halos as reported in  \citet{2001A&A...379.1052K} in the upper photosphere/lower chromosphere.
This effect of transients should be present in the active region power halos as well even though it may be less important by a more stable canopy and stronger wave signal. The power maps are strongly affected close to the network regions where jets are occurring ubiquitously. In the context of EUV coronal bright points  \citet{2015ApJ...806..172S} have demonstrated that the quasi-periodic oscillation in transition regions and corona above a network regions are due to repeated occurrences of jets around the network regions.

Similarly to the power halos, the magnetic shadow seen closer to the line core (H$\alpha+$0.543 and H$\alpha+$ 0.362 \r{A}) in the 3-minute power band can be affected by  the lifetime and distribution of the mottles. 
It is generally seen that, close to network regions, power above 5 minutes dominates whereas, 
in the inter-network regions, the dominant period is 
3 minutes \citep{1984A&A...130..331D,1990A&A...228..506D,1994ApJ...423L..67B,2000A&A...357.1093C,2001A&A...379.1052K,2009A&A...493..217T,2013SoPh..282...67G,2014MNRAS.443.1267B}. 
The H$\alpha$ core-intensity signal (see Figure~\ref{period_map}) is mostly dominated by $\ge$5-minute oscillations over the entire FOV,
whereas the Doppler velocity signal (see Figure~\ref{doppler}) shows a $\ge$5 minute dominant-period very close to the network region and 3-minute in the inter-network region. Similar behavior was also found by \citet{2007ASPC..368...65D}. 
The reason for this could be that, close to the network center, when the mottles travel 
upward, we observe blue shifts from material flowing towards the observer, 
but when these reach the magnetic canopy region, we will not be able to observe any LOS Doppler shifts as the material is flowing horizontally with a quasi-periodicity (the intensity fluctuations can still be observed). 
Rather the 3-minute shocks \citep{1992ApJ...397L..59C,1997ApJ...481..500C},
buffeting the canopy from below in the internetwork region, are observed in the Doppler signal. Hence, the lifetime of mottles won't affect the Doppler power map. 
The high intensity fluctuations produced by the appearance and disappearance of mottles cause more power at longer periods, instead of 
at 3 minutes.
We should point out that \citet{2014A&A...567A..62K} conjectured that the nature of the 7-minute power at
the chromospheric heights is not acoustic in nature.  
Using our simplistic model we tried to mimic the chromospheric power distribution and 
we find that the suppression of power in the 3 minute period band due to sudden fluctuations (like mottles), is only few percent (2--6\%) whereas the same fluctuations can highly influence longer-period (5--9 minutes) power.
Our analysis indicates that the observed long-period oscillation in the H$\alpha$ core and close to the network in  H$\alpha$ + 0.362 and H$\alpha$ + 0.543 \r{A} (see Figure~\ref{period_map}) 
arises due to longer lifetime of the mottles in the quiet-Sun network regions. 
From our observations we find that the magnetic shadow regions (network) show 60-70\% power reduction compared to the internetwork regions.
So, we conclude that although the power can be affected by the lifetime of
the mottles, the power suppression due to mottles may not be significant in the 3-minute. 
Hence, we conjecture that wave mode conversion may play a key role in forming magnetic shadows in the 3-minute power band. The slow waves may transfer part of their energy upon reaching the canopy layer and convert to fast magneto-acoustic modes. 
Due to high velocity gradients the fast mode generally reflects back and form magnetic shadow \citep{2006ApJ...653..739K,2006MNRAS.372..551S}.
In addition to this process, \citet{2016ApJ...817...45R} found that 
fast-to-Alfv\'en wave mode conversion may play an important role in this process and the
the fast wave energy can be converted to transverse Alfv\'en waves along the field lines. 
We should also point out that most of the theoretical work on the magnetic portals have not included the non-LTE effects which may 
play an important role in the coupled chromosphere where radiation effects are also important.

\section{Conclusions}
We studied the oscillatory behavior of the quiet Sun using H$\alpha$ observation encompassing network bright points.
The power-maps at different layers display 
the well known ``magnetic shadow'' and ``power halo'' features. Previously, these phenomena were interpreted in terms of acoustic waves interacting with inclined magnetic fields. We show that, power-maps in general, can be strongly affected by the lifetimes of these transients.
We propose that transients like RBEs, that occur ubiquitously in the solar atmosphere, can have a major effect on the formation of power halos in the quiet Sun.  
For magnetic shadows in the 3 minute band, the mode conversion seems to be most effective, whereas the power at longer periods is highly influenced by the presence of mottles. 
We should point out that the shorter time length of the time series also will have some effect on the power analysis.  
A very long time series should ideally be used for such purposes but high quality ground-based observations are rarely available for prolonged periods.
Most of the previous low cadence observations and numerical simulations have ignored the effects of small-scale transients while explaining the magnetic portal. Our high cadence observations reveal clear presence of these transients and thus
waves and transients may simultaneously be present within these structures and can collectively cause
the power enhancements and suppression. It will be very difficult to isolate and decouple these effects, although the dominant source for the formation of power halos appears to be the transients from our observation. We hope to quantify the contributions from these two sources in our future work, while studying the phase relation between intensity and velocity at different layers. With high spatial and temporal resolution observations we find that the quite sun chromosphere is highly dynamic where flows, waves and shocks manifest in presence of magnetic field to form an often non-linear magnetohydrodynamic system and future simulations should include all these effects. 
\acknowledgments
We thank the anonymous referee for his/her valuable comments which has enabled to improve the presentation. 
The authors would like to thank I. Kontogiannis for a helpful discussion and suggestions. 
This work was supported by UKIERI trilateral research grant of The British Council. The Swedish 1-m Solar Telescope is operated on the island of La Palma by the Institute for Solar Physics (ISP) of Stockholm University
in the Spanish Observatorio del Roque de los Muchachos of the Instituto de Astrof\'isica de Canarias. 
This research was supported by
the SOLARNET project {\color{blue}{(www.solarnet-east.eu)}}, funded by the European Commissions FP7
Capacities Program under the Grant Agreement 312495.
D.B.J. thanks STFC for an Ernest Rutherford Fellowship in addition to a dedicated standard grant that allowed this project to be undertaken.
\bibliographystyle{apj}

\end{document}